# Ultra-fast Vacancy Migration: A Novel Approach for Synthesizing Sub-10 nm Crystalline Transition Metal Dichalcogenide Nanocrystals

Pawan Kumar[1,2,3], Jiazheng Chen[1], Andrew C. Meng[2,4], Wei-Chang D. Yang[5], Surendra B. Anantharaman[1,6], James P. Horwath[2,7], Juan C. Idrobo[8,9], Himani Mishra[10], Yuanyue Liu[10], Albert V. Davydov[5], Eric A. Stach[2]* and Deep Jariwala[1]*

[1]Electrical and Systems Engineering, University of Pennsylvania, Philadelphia, PA, USA
[2]Materials Science and Engineering, University of Pennsylvania, Philadelphia, PA, USA
[3]Inter-university Microelectronics Center (IMEC), Leuven, Belgium
[4]Department of Physics and Astronomy University of Missouri, Columbia, USA
[5]National Institute of Standards and Technology, Gaithersburg, MD, USA
[6]Low-dimensional Semiconductors Lab, Metallurgical and Materials Engineering, Indian Institute of Technology-Madras, Tamilnadu, India
[7]Argonne National Laboratory, Illinois, USA
[8]Center for Nanophase Materials Sciences, Oak Ridge National Laboratory, Oakridge, USA
[9]Materials Science & Engineering, University of Washington, Seattle, USA
[10]Department of Mechanical Engineering and Texas Materials Institute, University of Texas, Austin, USA

**Abstract:**

Two-dimensional materials, such as transition metal dichalcogenides (TMDCs), have the potential to revolutionize the field of electronics and photonics due to their unique physical and structural properties. This research presents a novel method for synthesizing crystalline TMDCs crystals with < 10 nm size using ultra-fast migration of vacancies at elevated temperatures. Through *in-situ* and *ex-situ* processing and using atomic-level characterization techniques, we analyze the shape, size, crystallinity, composition, and strain distribution of these nanocrystals. These nanocrystals exhibit electronic structure signatures that differ from the 2D bulk i.e., uniform mono and multilayers. Further, our *in-situ*, vacuum-based synthesis technique allows observation and comparison of defect and phase evolution in these crystals formed under van der Waals heterostructure confinement versus unconfined conditions. Overall, this research demonstrates a solid-state route to synthesizing uniform nanocrystals of TMDCs and lays the foundation for materials science in confined 2D spaces under extreme conditions.

# Introduction:

Research into the one-dimensional (1D) and zero-dimensional (0D) confinement of two-dimensional (2D) materials has been ongoing for some time. Various successful approaches have been employed, such as electrostatic confinement, defect creation, modulation of band gaps through composition in lateral heterostructures, and strain-induced band gap modulation.[1, 2] However, crafting uniform, sub-10 nm crystalline structures of transition metal dichalcogenides (TMDCs) using a top-down process continues to pose a significant challenge. To date, TMDCs-based nanoparticles have been synthesized through either top-down or bottom-up approaches. Bottom-up approaches have utilized precursor concentration control as well as biomineralization approaches to form quantum dots (QDs),[3] while the top-down approaches exploit electrostatic confinement via nanofabrication of metal gates on 2D layers or decomposition from bulk TMDC crystals under sonication in the presence of surfactants.[4] Among the existing bottom-up techniques, maintaining control over the particles' crystallinity, size, and defects can be difficult. Similarly, the top-down techniques that are currently available still face the ongoing challenge of ensuring scalability and managing the control of phase and defects.

In this study, we present a unique approach to fabricating 2 nm to 15 nm lateral size particles through non-equilibrium thermolysis of 2D materials, resulting in the formation of nanocrystals that maintain the crystallinity of the starting single crystalline flake. *In-situ* and *ex-situ* processing, along with atomic-level characterization techniques such as aberration-corrected scanning transmission electron micro-spectroscopy and scanning probe micro-spectroscopy were used to analyze the shape, size, crystallinity, composition, and strain distribution of these nanocrystals. The electronic structure signatures of these nanocrystals, which differ from those of bulk materials, were also studied using near-field photoluminescence (PL), cathodoluminescence (CL), and electron energy loss spectroscopy (EELS). Our vacuum-based synthesis technique allows for the observation and comparison of defect and phase evolution under nanoscopic confinement vs open conditions. Overall, this study presents a solid-state route to synthesizing uniform nanocrystals of TMDCs and lays the foundation for materials science in confined 2D spaces under extreme conditions.

## Results and Discussion:

We demonstrate a novel method for fabricating TMDCs nanocrystals via a thermolysis process[5, 6], in which an ultrafast increase in temperature (< few seconds) leads to the rapid formation of sub-10 nm particles. We have studied this transformation both *in-situ*, using a micro-heater fabricated on a micro electro-mechanical system (MEMS) transmission electron microscopy (TEM) chip, and *ex-situ* in a tube furnace (see Methods for additional details).

Figure 1a shows an atomic model of the transformation of a continuous monolayer or few-layer TMDC into tiny islands through rapid thermolysis. For *in-situ* rapid thermolysis, the TMDC layer was transferred onto a micro-heater fabricated on a MEMS TEM chip (see Figure S1). To ensure uniform heat distribution to the TMDCs layer, a highly thermally conducting h-BN support layer was utilized (see Figure S2). Micro-holes were made in the SiNx membrane of the TEM chip using a focused ion-beam (FIB), allowing for atomic resolution imaging of the 2D layers (see Figure S1e and S1f). Figure 1b shows a high-angle annular dark field (HAADF) scanning transmission electron microscopy (STEM) image of $MoS_2$ at room temperature, consisting of monolayer and few-layer structures, as highlighted in the image. After thermal processing, the formation of randomly distributed tiny islands can be seen in Figure 1c. The thickness-dependence of this transformation is shown in the HAADF-STEM image, and two different particle size distributions are observed (see Figure S3 and S4 for details). Image segmentation and nanoparticle measurement details are described in the Methods. Figures 1d and 1e show STEM micrographs corresponding to the thermolysis of monolayer and few-layer $MoS_2$, respectively. The corresponding size distribution histograms are shown in Figures 1f and 1g, with the fitting curve indicating that monolayer $MoS_2$ transformed into multi-layered particles with a lateral size of approximately 4 nm to 6 nm, while few-layer $MoS_2$ transformed into particles with a size of approximately 10 nm. The particle size distributions fit approximately a Gaussian distribution with a mean of approximately 5 nm for the monolayer starting material, while a Lorentzian distribution curve provides a better description of the few-layer starting material, with a median of ≈10 nm. The ability to control the particle size through control of the initial layer thickness allows for the design of various engineered nanostructures. The shape distribution analysis is also shown in the inset of Figures 1f and 1g (histogram view for area/perimeter$^2$) and in SI Figure S5, where a tendency to form two different shapes (with hexagon being the dominant shape) can be seen. Further, an atomically resolved HAADF-STEM image is shown in Figure 1h which formed after monolayer disintegration. We also analyze these nanocrystals in phase contrast TEM image, as shown in Figure 1i. The FFT patterns extracted from the respective STEM and HR-TEM images (a rectangular box in a different color) are also shown in Figure 1j, highlighting the preserved hexagonal crystal symmetry after thermal disintegration. The FFT patterns presented as insets show that crystal orientations of different nanoparticles are not aligned with respect to each other. The other brighter spots (aside from the six streaks) marked rectangular in the FFT pattern correspond to the crystalline h-BN layer.

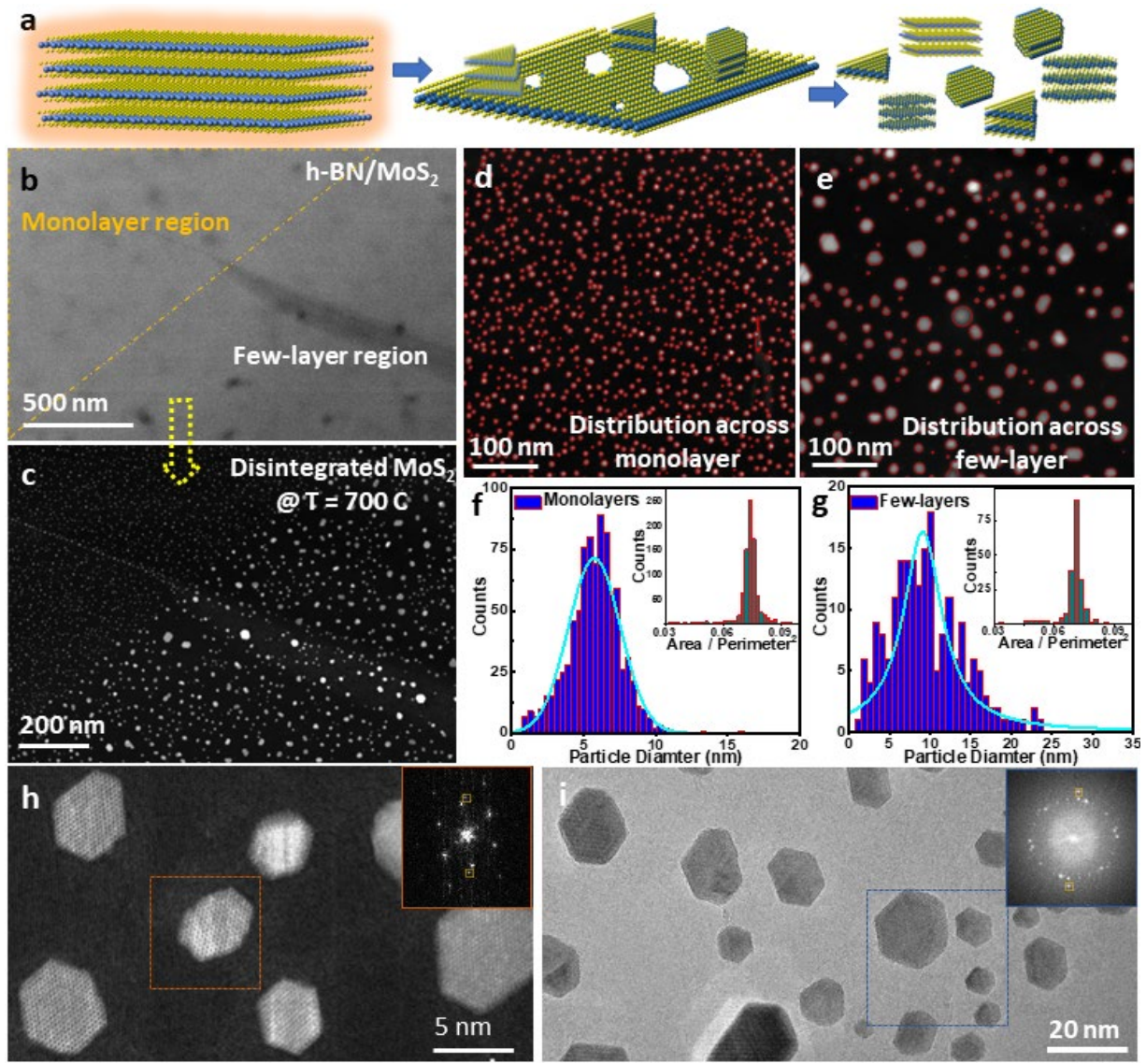

***Figure 1:*** (a) An atomic model illustrating the thermolysis process for the formation of crystalline sub-10 nm particles of $MoS_2$ of varying shapes. (b) Dark-field STEM image of a continuous monolayer and few-layer $MoS_2$ supported with h-BN pre-rapid heating and thermolysis, and corresponding (c) thermolysis-induced disintegration into crystalline nanoparticles from the parent continuous layer. (d, e) Magnified view of the disintegrated nanoparticles from the monolayer and few-layer regions which were used for the size and shape analysis. The particle size statistics were analyzed using convolutional neural network (CNN)-based machine learning as presented in Figures f and g for monolayer and few-

layer regions, respectively. (h) Atomic resolution HAADF-STEM image of the disintegrated nanoparticles made from a monolayer region post thermolysis. (i) phase-contrast TEM image of the nanoparticles disintegrated from the few-layer region; Corresponding FFT patterns extracted from the marked rectangular box regions in Figure h & i respectively.

The formation of highly confined multi-layered particles was also observed for other TMDCs such as MoSe$_2$, WSe$_2$, and WS$_2$. The as-formed crystalline particles from various TMDCs are shown in bright-field STEM images in Figure 2a-c, respectively, at different thermolysis temperatures. The inset, lower magnification images provide a larger area view of the *in-situ* transformed crystalline particle formation (see Figures S6 and S7 for more details). Selenium-based TMDCs mostly form connected chains of islands after thermolysis, while sulfur-based TMDCs form isolated particles. The decomposition of chalcogenides in a selenium-rich environment at raised temperatures is known to precipitate an amorphous-like (fine-grain) layer consisting of selenium and carbon surrounding the larger crystallite (TMDC islands).[7] The amorphous-like substance may tether the selenium-based TMDCs that experience thermolysis to form chains of islands.

While performing the *in-situ* thermolysis process, a significant challenge is ensuring the successful and clean transfer of an atomic thin 2D layer over the SiNx membrane due to polymer contamination. The dry transfer technique, which is the most feasible and widely used method for transferring multiple layers one after another, almost always leaves a thin polydimethylsiloxane (PDMS) residual layer (see the Method section for details) even when performed in a glove box after fresh exfoliation. This thin polymer layer forms agglomerated islands after annealing treatment (annealing at 300 °C under a forming gas environment for 8 h to 12 h) following each layer transfer process. We observe that the PDMS residue at the interface of the h-BN and TMDCs layers remaining from the stamp transfer process acts as a barrier layer, resulting in non-uniform heat spread during rapid thermolysis (see Figure S8). This non-uniformity can be eliminated by using the nano-compression technique[8] before the thermolysis process. The atomic force microscopy (AFM) height image of the transferred h-BN/WSe$_2$ layer on the TEM grid before and after (Figure 2d) the nano-compression process clearly shows the flatness of the heterostructures (here, h-BN/WSe$_2$). The trapped PDMS layer was removed from the interface using a sharp AFM tip to improve adhesion.[8, 9] The microscopically flat and better-interfaced heterostructure (h-BN/WSe$_2$) layer leads to the uniform and complete transformation of the WSe$_2$ layer into sub-10 nm size islands compared to the non-compressed regions, as seen in Figure 2e. These compressed, flat regions are clearer to image in the electron-transparent region (visible bean shape regions in figure 2d) available in the TEM grid where all the *in-situ* imaging was performed. In principle, the non-compressed regions can also be transformed into nanoparticles if the thermolysis time is increased to more than 10 sec. However, the stoichiometry of the TMDCs layer is lost, and metal nanocrystals begin to nucleate due to excessive loss of the chalcogen instead of leading to stoichiometric TMDCs nanoparticle formation (see metal crystal formation in Figure S9 d-f).

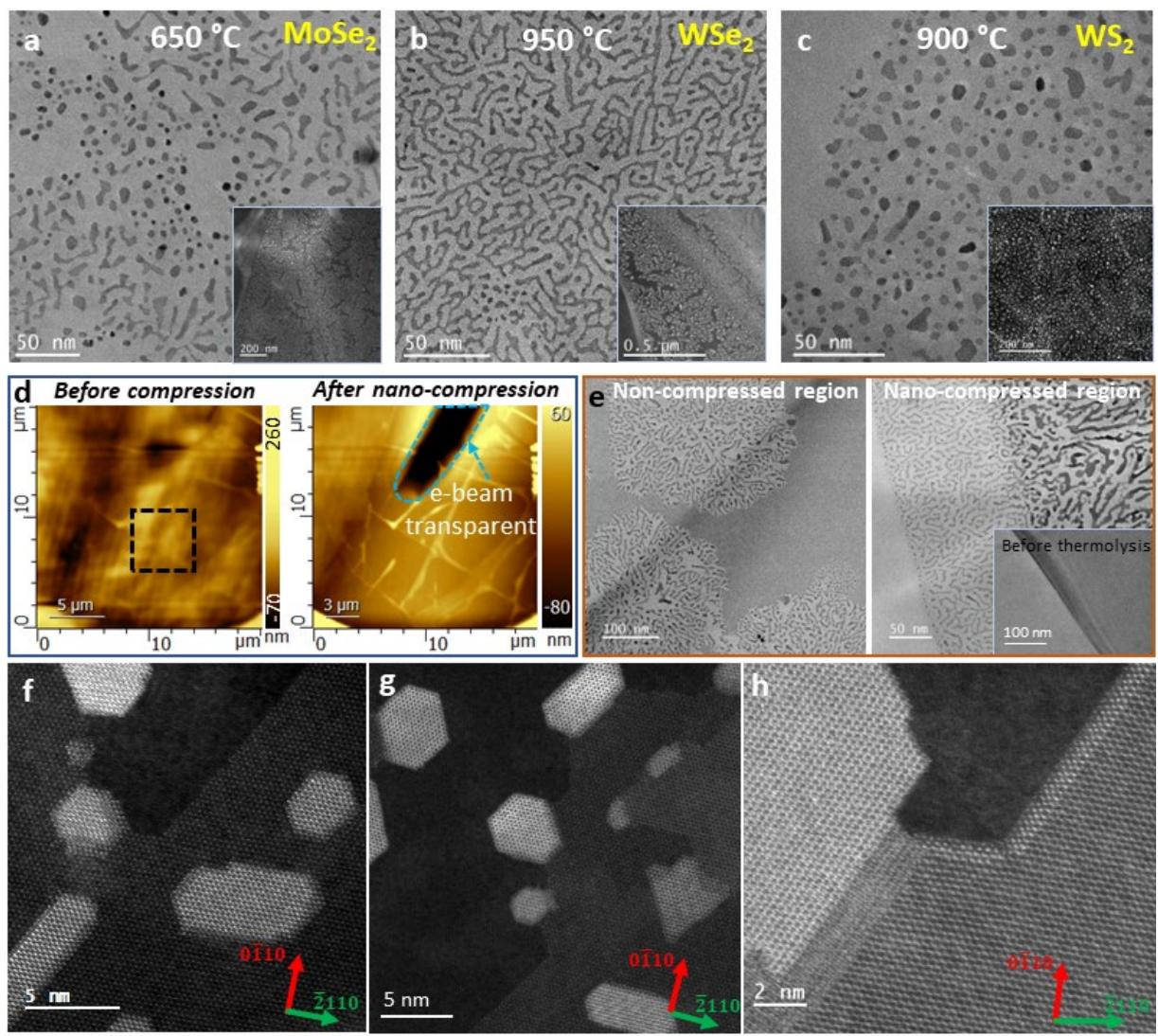

*Figure 2:* Bright-field TEM images of the (a) $MoSe_2$, (b) $WSe_2$ and (c) $WS_2$ particles after thermolysis at respective temperatures. The inset shows a lower magnification view of the transformed particles. (d) AFM height images of the transferred continuous $WSe_2$ layer before and after compression at the pre-heating stage. Nano-compression was performed using an AFM tip in contact mode (applying 1μN force) before height scanning. Corresponding (e) bright-field TEM images of the regions in (d) after rapid-heating show how nano-compression forms better interfaces. (f, g) An atomically resolved HAADF-STEM image of the disintegrated particles showing layer-by-layer disintegration during thermolysis. (h) Atomic resolution HAADF-STEM image showing edge reconstruction, which was observed at some locations.

Non-equilibrium thermolysis relies on the rapid migration of vacancies, primarily S-vacancies in the case of $MoS_2$ or $WS_2$, to synthesize TMDCs nanoparticles.[6] While it can be challenging to observe the complete formation process in detail, we were able to confirm the different stages of the formation using atomic resolution HAADF-STEM images as shown in Figure 2f, g (also shown in Figure S10-S12). These images show layer-by-layer diffusion and the formation of nanoparticles in few-layer TMDCs. We also observed edge reconstruction and phase transformation at some sites during the thermolysis process (Figure 2h). The edge reconstruction is mostly visible in cases where the temperature rise occurred at a slower rate than intended in the experiment. The corresponding FFT (see Fig S10, S11) confirms the similar orientation of the nanoparticles and underneath layer. Further, we characterized the nanoparticles using a range of tools, including AFM, STEM-EDS (energy dispersive X-ray spectroscopy), and 4D-STEM mapping. Figure 3a shows an AFM height image, with corresponding line profiles shown in Figure 3b. The particles or constrictions (nanoribbons) have thicknesses ranging from a bilayer ($\approx$ 1.7 nm) to a few layers ($\approx$ 3 nm), and lateral dimensions ranging from 3 nm to 10 nm. STEM-based EDS mapping, shown in Figure 3c, was used to analyze the elemental composition of the particles. The STEM image of the survey area and the marked green rectangle region show a uniform distribution of Mo and S elements (see Figure S13 for additional EDS images taken from a batch of samples). It is worth noting that EDS mapping at 200 kV may result in the loss of sulfur atoms due to knock-on damage during spectral imaging.[10] Relative crystallographic orientation changes can occur between nanoparticles formed from the $MoS_2$ single-crystal flakes (see Figure S14). 4D-STEM is used to perform differential phase contrast imaging (Figure 3d, #2) to characterize strain (Figure 3d, #3-5) in the $MoS_2$ nanoparticles. Differential phase contrast (DPC) mapping is measured using the center of mass method, which is influenced by electromagnetic fields in the sample as well as sample crystallographic orientation.[11] Figure 3d (#2) indicates slight DPC in the $MoS_2$ particles: this can arise either due to an electromagnetic field in the material, strain differences, crystallographic orientation differences, or a combination of all three. The strain maps show that the lattice parameter of the $MoS_2$ nanoparticles is relatively uniform and varies on the order of 1 %. Assuming there are no electromagnetic fields in the material, this would suggest that contrast in the DPC image arises mainly due to small differences in crystallographic orientation. This could be consistent with our previous observations of variations of tilt in neighboring crystals of $PtSe_2$ formed by thermolysis.[12]

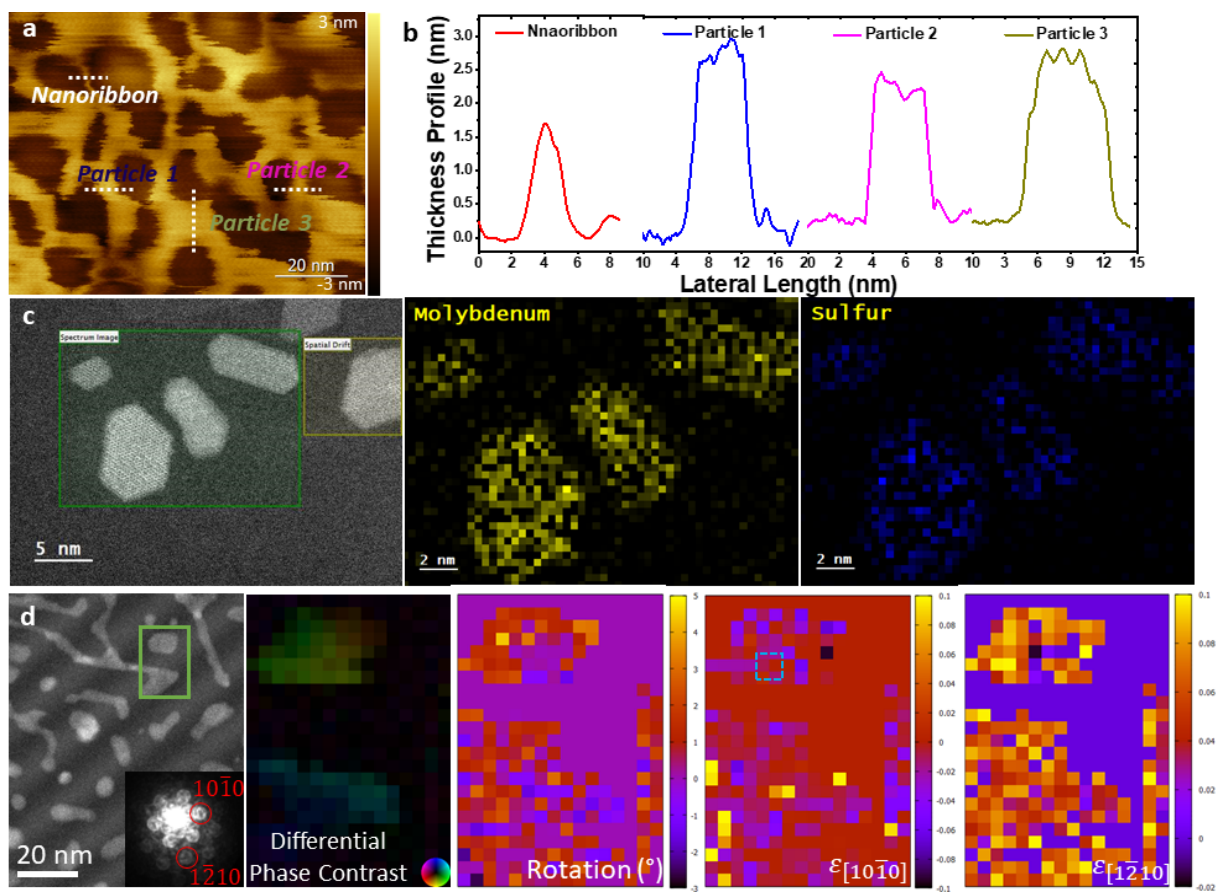

***Figure 3:*** (a, b) Atomic force microscopic height (topography) image of the disintegrated nanoparticles/nano-constrictions and the corresponding line profile analysis. the data shows clear formation of 4 nm to 10 nm wide multi-layered (and up to 3 nm thick) particles and constrictions. (c) EDS elemental mapping ($MoS_2$) confirms the formation of stoichiometric nanoparticles. (d) 4D-STEM analysis of the disintegrated $MoS_2$ nanoparticles showing (left to right) dark field STEM image, differential phase contrast (DPC) image, rotation and strain maps in two component directions, $[10\bar{1}0]$ and $[1\bar{2}10]$.

Non-equilibrium thermolysis leads to several phenomena including vacancy creation and migration (especially S-vacancies), structural reconstruction, phase changes, and stoichiometric variations. Here, we used a confined configuration of 2D materials undergoing rapid thermolysis to more closely examine these phenomena that occur during non-equilibrium thermolysis. To create this configuration, we added a protective h-BN layer on top of the h-BN/$MoS_2$ configuration (shown thus far in Figures 1-3), thereby completely enclosing the $MoS_2$ flake (as shown in the atomic model in Figure 4a). The resulting encapsulated configuration is also shown in Figure S15 in supporting information. Rapid-heating was then performed for inducing non-equilibrium thermolysis on this $MoS_2$ flake, which was encapsulated by a few-layer h-BN from both the top and bottom (h-BN/$MoS_2$/h-BN) creating a closed "sub-picolitre reactor" system. The dark-field STEM image in Figure 4b shows that the disintegrated $MoS_2$ contains both monolayer and few-layer regions (marked with two different colors). A magnified, high-resolution HAADF-STEM image of the monolayer region (Figure 4c) reveals the formation of nanoparticles in pockets, similar to the non-confined h-BN/MoS$_2$ configuration. The encapsulated system experiences additional pressure from the top and bottom few-layer h-BN, which affects the thermolysis process. This increased pressure causes the temperature at which the nanoparticles form to increase (see Figure S16-S18). This is because the log of the diffusion coefficient decreases as a function of pressure proportional to the activation volume,[13] raising the temperature at which the transformation occurs. The closed "sub-picolitre reactor" also helps maintain the stoichiometry of the formed nanoparticles since they prevent chalcogen escape. Further, by increasing the duration of the peak thermolysis process time, we observe complete thermolysis in the encapsulated region, forming crystalline particles while simultaneously also observing that the adjacent unencapsulated monolayer becomes thicker and turns completely metallic due to chalcogen sublimation (see SI Figure S17 c-e and S18). In contrast to the monolayer regions, we observed unique occurrences in the case of few-layer regions (as shown in Figure 4d). A high-magnification HAADF-STEM image shows the formation of faceted trenches with mostly triangular or trapezoidal shapes (see SI Figure S19). The preferentially oriented trench formation (see Figure 4d inset) under complete confinement suggests that line defects are likely contributing into the edge restructuring process. Previous studies have identified various types of edges in 2D TMDC layers after high-temperature treatment.[14] Figure 4e shows the atomically resolved structural arrangement of bundled edges on one side of the trenches. The other sides of the trenches appear normal, with no atomic accumulation. The unique edge restructuring observed in this study can be attributed to the confinement in the z-direction, which prevents atoms from migrating from one layer to another in a few-layer 2D system, leading to the formation of edge bundling. This unique phenomenon merits further detailed investigation, but is beyond the scope of this manuscript.

To further evaluate the shape stability of the disintegrated particles, we used Density functional theory, DFT (Nudged Elastic Band Method) to analyze the different shapes obtained. We found that the non-confined configuration tended to form circular or hexagonal shapes, while the confined and encapsulated configurations resulted in slower, stoichiometric transformation and the formation of thermodynamically stable, triangular shapes in the case of 3-fold symmetric MoS$_2$. Considering the nanoparticle formation process, there are two main driving forces for nanoparticle shape evolution: 1) diffusion of atoms at the particle surface/edge, and 2) solid-state reaction of TMDC decomposition at the surface/edge. Under diffusion-limited kinetics, the nanoparticle shape will be determined by the diffusion coefficient anisotropy, whereas under reaction rate-limited kinetics, the nanoparticle shape will be determined by the anisotropy of the TMDC's decomposition reaction. Figure 4f shows the simulated diffusion coefficient anisotropy for a number of edge geometries. The results are consistent with diffusion-limited kinetics in the confined geometry. On the other hand, nanoparticles formed in the non-confined geometry exhibit very little anisotropy. We can rationalize this by considering the effect of the h-BN encapsulation as an effective pressure increase, which decreases the diffusion coefficient and results in diffusion-limited shape evolution. If diffusion is relatively fast compared to evaporation, the shape will be closer to the thermodynamic equilibrium shape (a triangle due to the 3-fold symmetry of $MoS_2$). If not, the shape will be closer to the kinetic shape (more round or dendritic). Fast diffusion along the edge is possible at high temperatures (with an energy barrier of 0.8 eV), leading to shape changes that are closer to the thermodynamic equilibrium shape (Figure 4f). Intermediate energy states for sulfur diffusion at the $MoS_2$ edge are shown in SI Figure S20-21.

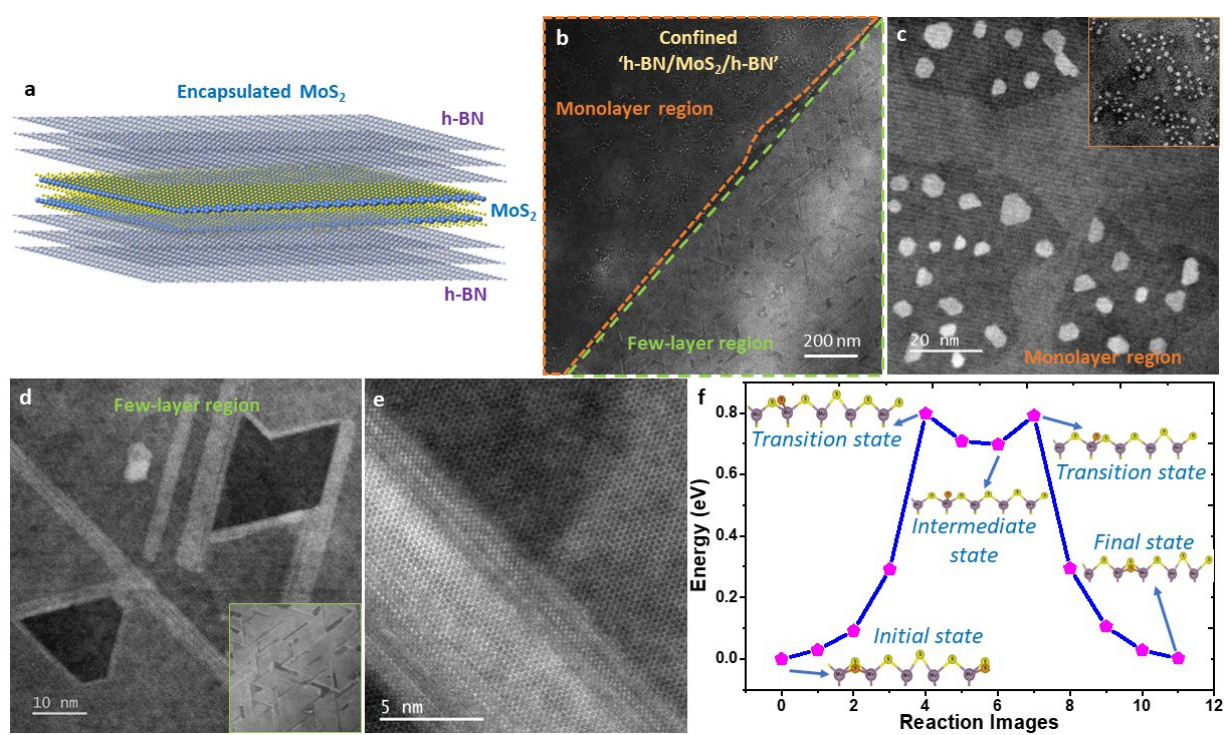

*Figure 4:* (a) Schematic showing the encapsulated 2D $MX_2$ TMDC layer in between the top and bottom few-layer h-BN. (b) The dark-field STEM image of the disintegrated particles under 2D encapsulation highlights two different thicknesses as outlines by the orange and green dashed boundaries. The corresponding (c) high-resolution STEM image for the monolayer region shows pockets of disintegration with triangular and truncated hexagonal particles. (d) STEM image for the few-layer region, showing unique edge reconstruction and faceted-trench formation (see inset) unlike the monolayer case which disintegrated into geometric ≈5 nm size particles under encapsulation. (e) Atomically resolved HAADF-STEM image of one of the as-formed restructured trenches showing the armchair-type edge atomic structure and the surrounding continuous region that has not disintegrated. (f) First-principles calculation of sulfur diffusion at the MoS$_2$ edge and the corresponding energy explain the tendency of the differing shapes based upon stable edge formations under encapsulation.

Investigating the optical characteristics of the nanoparticles produced in our study was challenging using the *in-situ* MEMS grid. The free-standing SiNx membrane has overlapping background emission spectra (likely from defects and its non-stoichiometry) and in addition, its non-reflective surfaces do not allow for sufficient signal to be distinguished between the emitting nanoparticles and the background when using micro-PL spectroscopy. The near-field spectral luminescence measurement was also unsuccessful due to the free-standing SiNx membrane interaction with the gold-coated near-field tip. We optimized the thermolysis process using an ex-situ approach to address this issue and develop a more scalable method for producing nanoparticles (see the Method section for more details). The *ex-situ* approach, which involves rapid thermal annealing (RTA), produced the identical quantum particles and allowed for their characterization using both far-field and near-field micro-PL spectroscopy. Figures 5a and 5b show AFM height images of a MoS$_2$ monolayer (includes partly bilayer as well) before and after being heated at 800 °C for 45 sec (inside a customized furnace-based RTA system), showing the topography before and after *ex-situ* thermolysis. A noticeable structural evolution is visible throughout the MoS$_2$ flake when comparing the two images. Finding the optimal conditions for producing nanoparticles is challenging as the formation of defect-mediated intact nanoparticles requires a narrow window, with either too many defects leading to particle destruction or slow evaporation leading to the formation of metal nanoclusters. Structural transformation is more pronounced at the monolayer edge and monolayer-bilayer interface due to the higher reactivity of these sites as predicted by theoretical energy calculations.[15] Additionally, the bright dots in the AFM image are Mo clusters rather than MoS$_2$ nanoparticles in some parts of the flake due to the complete loss of sulfur atoms during the longer heating process. We adopted a slightly longer (45 sec) heating time to ensure complete transformation (see SI Table S1 and Figure S22 for complete optimization results). A complete transformation occurs in the monolayer case, while a few layers undergo thermolysis with the remaining underlying layer and nanoparticle formation under the same temperature and conditions.

The blue shift in the PL spectra of the monolayer and bilayer cases, shown in Figure 5c, suggests the formation of defect-mediated nanoparticle emission, consistent with previously reported quantum dots[16] while contrary with the emitters based on 2D materials.[2, 17] However, many areas of the flake show near-zero PL emission, particularly in areas that have undergone overheating (see SI Figure S22). To study the localized emission properties of the nanoparticles, we used near-field Tip-enhanced Photoluminescence spectroscopy (TEPL). This technique provides high resolution in terms of spatial variations and localized emission features from the active material region. Figure 5d shows the overlapped spatially mapped TEPL signal of the transformed regions after *ex-situ* thermolysis with AFM topography image. A comparison of the TEPL spectral response from a few spots in the center of the flake (Figure 5e) shows no significant peak shift in relation to the far-field PL (Figure 5c). A quenched PL signal is observed towards the flake edges suggesting the formation of thicker, multilayer particles that appear bright in AFM topography images. More transformation (conversion into metallic Mo particles) at the edges compared to the center positions under the optimized *ex-situ* RTA processing. However, further studies and optimization may be necessary to fine-tune the controlled nanoparticle formation and to simultaneously avoid the formation of bulk metal particles (also SI Figure S23 for MoSe$_2$).

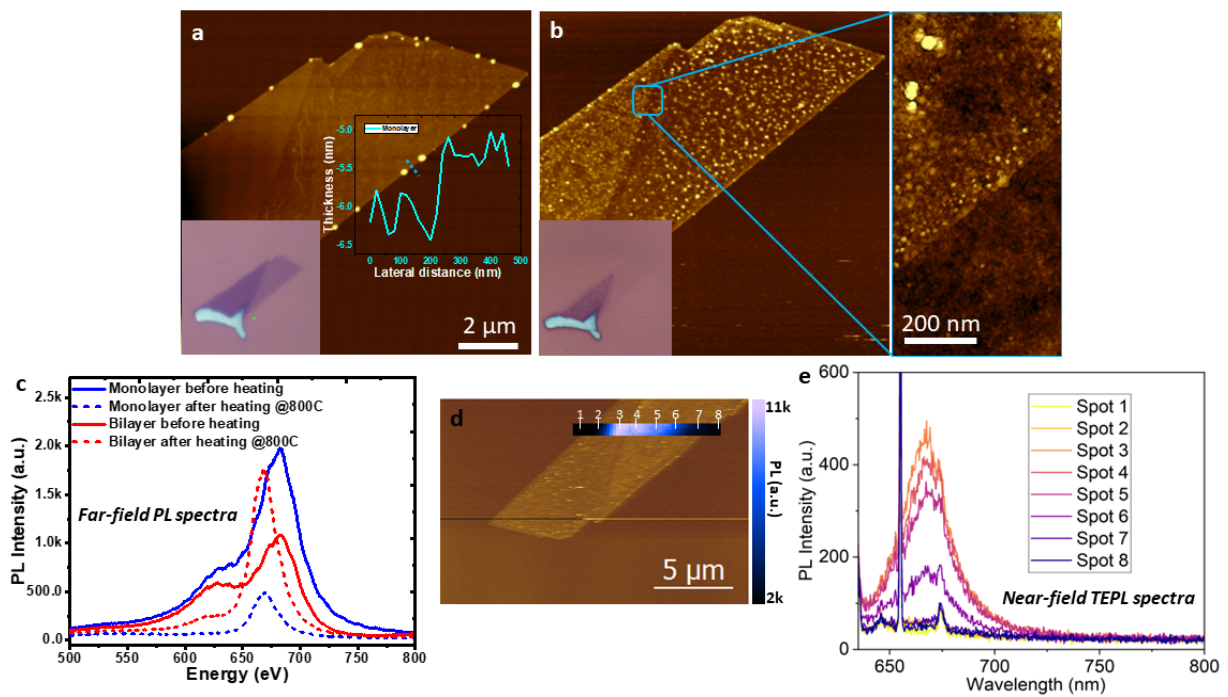

*Figure 5:* (a, b) AFM height image of the MoS$_2$ flake before and after the *ex-situ* rapid thermal heating at a temperature of 800 °C for 45 s respectively. Insets show the respective optical micrographs. The high magnification AFM height image is also shown on the right side of (b) showing the morphology of both the monolayer (bottom section) as well as the bilayer (top section) after thermolysis. (c) Far-field PL spectra have been plotted for the *ex-situ* thermolysis (pre- and post-heating at 800 °C for 45 s) where we can see the energy shift along with an increase in the emission intensity. (d) Near-field tip-enhanced PL spectroscopy is carried out to measure emission characteristics of as-formed nanoparticles and locally attached a spatially resolved TEPL spectral map. (g) Point spectra from several points in the TEPL map from the high-intensity region in the center shown in (f).

To understand the defect-mediated emission of MoS$_2$ after the thermolysis, we probe the excitation of valence electrons using electron energy loss spectroscopy (EELS) in a state-of-the-art probe-corrected STEM with a highly coherent and monochromatic electron source. The electron energy losses associated with A, B, and C excitons of MoS$_2$ are identified in both the confined nanoparticle and the film after the thermolysis, in reference to the simultaneously measured h-BN polariton peak position (Figure 6 a and b, SI Figure S24). Although the post-thermolysis intensity attenuates (pre-thermolysis intensity not shown here), the EELS result suggests a similar light extinction property before and after the thermolysis. The extinction of nanoparticles is known as a combined effect of energy absorption and scattering[18], which is sensitive to the particle's geometry and dielectric environment at the single-particle level[22]. Therefore, we adopt a multimodal approach to measure the electron-beam-excited scattering, i.e., cathodoluminescence (CL) that is spatially correlated with EELS in a TEM with a custom-built optical spectroscopy system[19]. For the monolayer MoS$_2$ encapsulated by h-BN before and after the thermolysis, we confirm that the EELS peaks of A, B, and C excitons (blue solid curves, Figure 6 c and d) still exist but are attenuated. Concurrently, the CL indicates a strong encapsulation-induced bulk excitonic emission (red solid curves, Figure 6d), while direct band edge emission is not present (SI Figure S25a). Enhanced C-exciton emission that is higher in energy, concurrently with a lack of primary A or B exciton emission implies that the highly confined, high-energy excitons resonating within the particles have insufficient time or phonon interactions to relax into a lower energy state, causing them to emit at this higher energy state. This agrees with our theoretical prediction using the boundary element method (SI Figure S26) that shows the quenched A and B excitons (green shadowed spectral range in Figure 6e) as well as the blue-shifted C exciton (blue arrow in Figure 6e) in CL emission from a MoS$_2$ disk supported by h-BN.[20] Similar emission characteristics are seen in the film region of Figure 6a (SI Figure S25) suggesting such region may mainly consist of nanoparticles with a background intact layer of MoS$_2$ resulting from incomplete thermolysis.

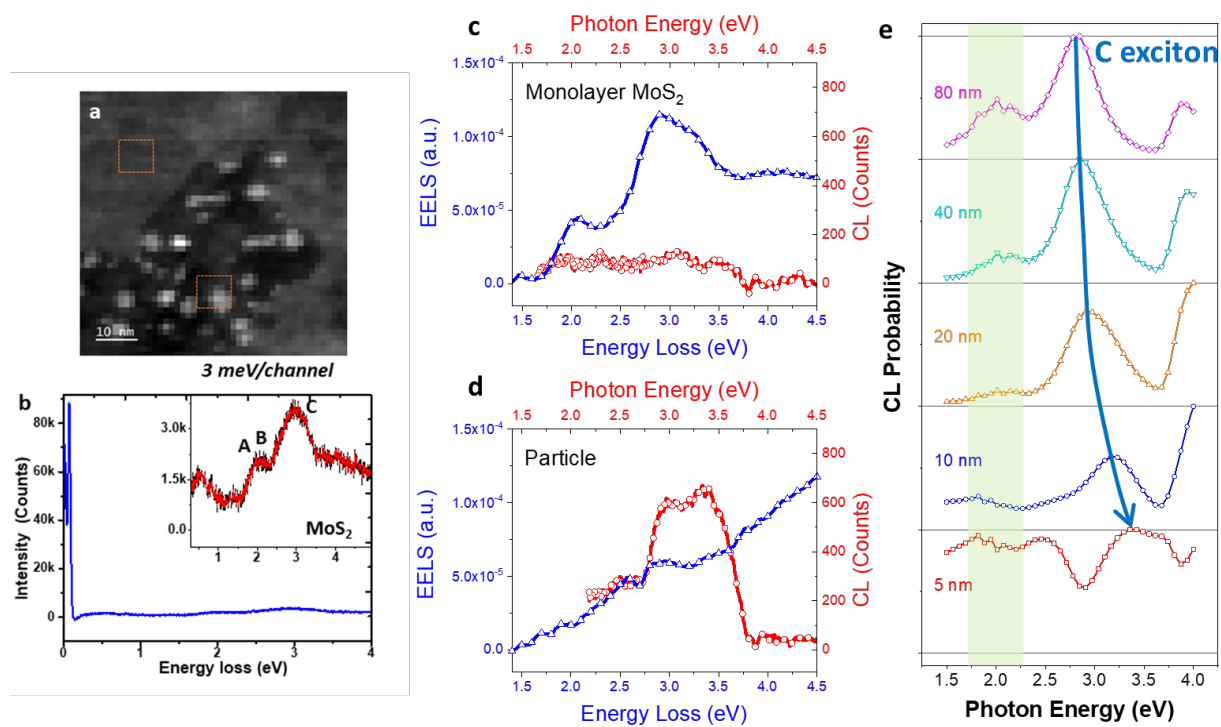

*Figure 6:* EELS and Cathodoluminescence (CL) analysis of the MoS$_2$ nanoparticles performed before and after thermolysis. (a) HAADF-STEM image of as-heated MoS$_2$ and (b) corresponding EELS spectral response was taken from the marked rectangular regions, inset shows the magnified view of the Valence-edge EELS spectra referring to A, B and C excitons of MoS$_2$ layer. (c) Spatially correlative EELS (blue, triangle) and CL (red, circle) spectra from the monolayer as MoS$_2$ before rapid thermolysis. (d) Post thermolysis of encapsulated monolayer MoS$_2$, spatially correlative EELS (blue, triangle) and CL (red, circle) spectra for the as-formed MoS$_2$ particle region. (e) Theoretical prediction calculated using the boundary element method for a MoS$_2$ disk (1 nm in thickness) with a diameter of 80 nm (magenta, rhombus), 40 nm (turquoise, downward triangle), 20 nm (orange, upward triangle), 10 nm (blue, circle), and 5 nm (red, square) supported by a h-BN substrate (1 nm in thickness).

## Conclusions:

In conclusion, we have demonstrated sub-10 nm uniform particle formation via thermolysis of single crystalline monolayer and few-layer TMDC flakes with h-BN supported as well as in an encapsulated van der Waals heterostructure. The h-BN supported TMDCs monolayer undergoes a uniform NPs formation having an average diameter ≈ 5 nm, while with the few-layer case, it is closer to ≈ 10 nm. The encapsulation is observed to suppress thermolysis temperature requirements and leads to the formation of highly crystalline and stoichiometric nanoparticles. The formed nanoparticles exhibit electronic signatures different from bulk 2D TMDCs, though no distinct signatures of quantum confinement were observed in either photoluminescence or cathodoluminescence spectroscopy. In summary, our work opens a new route to producing crystalline 0D and 1D nanostructures from 2D van der Waals layers and allows the study of phase transformation and diffusion phenomena in solid-state systems under highly confined and extreme environments.

## Materials and Methods:

**Materials Synthesis**

A mechanical exfoliation technique is used to make the monolayer as well as the few-layered TMDCs ($MoS_2$, $MoSe_2$, $WS_2$, $WSe_2$) and few-layered h-BN. After confirmation of the thin exfoliated layers using an optical microscope, it was then transferred to the target MEMS heating chip. The site-specific targeted transfer was achieved by the PDMS-based dry transfer technique[23]. All the transfer was precisely done to make sure the TMDC layers face free-standing regions to achieve atomic-resolution STEM microscopy. Once a TMDC or h-BN layer was transferred to the MEMS chip over located free-standing $SiN_x$ membrane area, it was then annealed in forming gas up to 300 °C for 12 h to 15 h to reduce the PDMS contaminants which left over during the dry transfer technique. Successively next TMDCs or h-BN layer was transferred further to synthesize the confined configuration like h-BN/$MoS_2$/h-BN.

**Materials Processing**

As received prefabricated heating chips (MEMS-based) were taken to the focus ion beam (FIB) chamber for fabricating holes in the TEM transmitting area. Very precise and site-specific sub-micron holes were made into $SiN_x$ membranes using the Xe-plasma-based FIB technique (TESCAN S8000X Focused Ion Beam / Scanning Electron Microscope). FIB drilling uses 5 pA current at 10 kV accelerating voltage to substantially reduce the redeposition process as well reduction of spatial contamination. The *in-situ* thermolysis process is achieved using a prefabricated MEMS heating chip loaded into an *in-situ* vacuum heating TEM holder (Hummingbird Scientific LLC, USA). The rate of heating was ultra-fast, and the target temperature was achieved in a few seconds at the rate of 100 °C/s, similarly, the cooling rate was achieved too. *Ex-situ* thermolysis was performed in a custom-made rapid thermal annealing system. The single-zone quartz-tube-based heating furnace (ThermoFisher Scientific, Lindberg Mini) was converted into a rapid thermal system. A target temperature (processing of 2D layer) was first maintained for a few minutes in a continuous Ar-flowing quartz tube (before, the Quartz tube is evacuated and flushed with Ar-gas multiple times using a rotary pump). A quartz rod with an attached sample holder was used to insert/withdraw quickly in/from the hot zone (where the target temperature was maintained) within a frame of seconds. Several processing was done to reach the as-optimized target temperature and thermolysis time to achieve quantum particle formation (see detail parameters in Table S1). Each optimization process needs several steps including Exfoliation of fresh TMDC layers, Dry transfer to $SiO_2$/Si substrate, Forming gas based annealing treatment, AFM and Optical microscopy characterization before and after thermolysis and micro-PL/Raman spectroscopic analysis.

**Materials characterization**

An *in-situ* TEM vacuum heating holder from Hummingbird Scientific LLC is used to perform all the *in-situ* thermolysis processes and analyses. Transmission electron microscopy (TEM) and scanning transmission electron microscopy (STEM) technique is used to do all the *in-situ* thermolysis processing. JEM 200 from JEOL microscope with 200 kV accelerating voltage has been utilized for all the TEM/STEM and 4D-STEM analyses. For 4D-STEM, data were collected using STEMx (Gatan) with a Gatan OneView camera (4k × 4k) using drift correction once every row. A 10 μm condenser aperture was used for 4D-STEM experiments to limit overlap between diffraction disks. 4D-STEM DPC mapping was performed by measuring beam displacement using the center-of-mass method. Gatan Microscopy Suit (GMS v3) was used to analyze the strain. To avoid any beam damage and irradiation effect, we keep changing the image acquisition positions as much as possible for every scan. Probe-corrected STEM (JEOL NEO ARM 200) is used to acquire atomic-resolved dark and bright-field STEM images. EDS mapping was done in the NEO-ARM 200 containing a high-speed dual detector which facilitates fast acquisition with high gain helpful for beam-sensitive material analysis. Further, valence band-edge EELS analysis is done using NION Ultra-STEM 100 system having a monochromatic electron source with an inverted e-gun configuration used to resolve the A, B and C excitons. The band-edge EELS spectra were referenced with respect to h-BN phonon lines. EELS is measured at 80 keV accelerating voltage and 0.75 pm probe size which attains 35 meV zero loss feature. Cathodoluminescence (CL) along with EELS spectra are recorded using an environmental TEM equipped with a monochomrated Schottky field-emission gun (FEG) at 80 kV accelerating voltage and ≈ 1 nm probe size, which obtain an energy spread of 80 meV. The spatially correlative measurement of EELS and CL is achieved using a custom-built optical spectroscopy system that inserts a parabolic mirror into the narrow gap between the sample holder and the lower polepiece of the objective lens.

Image segmentation and nanoparticle measurement were performed using a series of filters and morphological operations performed using the scikit-image python library.[21] HAADF-STEM images were initially processed using a gaussian filter, followed by morphological reconstruction using dilation and erosion. This produces a nearly binary version of the STEM image which can be finally segmented using an Otsu threshold. A connected components algorithm is used to link individually segmented pixels in particles, which can then be measured and analyzed.

**Theoretical modelling**

Here, we utilize climbing image nudged elastic band method[22] to calculate the energy barrier for S vacancy diffusion. The system is modelled by a ribbon with a S vacancy located at the edge (see the SI fig. S20 for the full structure). We perform DFT calculations using VASP software[23] with 400 eV energy cutoff, PBE functional and D3 method for vdW correction[24] and 3×3×1 k-grid. The initial state, final state, and other states in between are shown in Figure 4f. We see two saddle points and a local minimum (intermediate state) between these two saddle points. Taking this intermediate state as the final structure, we again perform NEB calculations with 10 additional states between the initial state and intermediate state. The energy profile of this reaction pathway shows a saddle point with a reaction barrier of 0.8 eV exactly at the middle as shown in supplementary Figure S21.

**Supporting Information**
SI contains the information of *in-situ* and *ex-situ* processing techniques; Details of MEMS heating TEM chip; Optical microscopic images of the fabrication of encapsulated configuration; TEM and STEM images for before and after *in-situ* heating of the h-BN/MoS$_2$ and h-BN/MoS$_2$/h-BN configurations; Edge reconstructions by atomically resolved HAADF-STEM images; Nanoscale EDS mapping; DFT calculations for the edge vacancies and formations energy; Near-field TEPL spectra for MoSe$_2$; Cathodoluminescence measurements and their fitting/analysis.

# Acknowledgement:

D.J., E.A.S. and P. K. acknowledge primary support from National Science Foundation (DMR-1905853) and support from University of Pennsylvania Materials Research Science and Engineering Center (MRSEC) (DMR-1720530) in addition to usage of MRSEC supported facilities. The sample fabrication, assembly and characterization were carried out at the Singh Center for Nanotechnology at the University of Pennsylvania which is supported by the National Science Foundation (NSF) National Nanotechnology Coordinated Infrastructure Program grant NNCI-1542153. D.J. also acknowledges partial support for this work by the Air Force Office of Scientific Research (AFOSR) FA2386-20-1-4074.
NIST Disclaimer: Certain equipment, instruments, software, or materials, commercial or non-commercial, are identified in this paper in order to specify the experimental procedure adequately. Such identification is not intended to imply recommendation or endorsement of any product or service by NIST, nor is it intended to imply that the materials or equipment identified are necessarily the best available for the purpose.

# Supporting Information

## Ultra-fast Vacancy Migration: A Novel Approach for Synthesizing Sub-10 nm Crystalline Transition Metal Dichalcogenide Nanocrystals

Pawan Kumar[1,2,3], Jiazheng Chen[1], Andrew C. Meng[2,4], Wei-Chang D. Yang[5], Surendra B. Anantharaman[1,6], James P. Horwath[2,7], Juan C. Idrobo[8,9], Himani Mishra[10], Yuanyue Liu[10], Albert V. Davydov[5], Eric A. Stach[2]* and Deep Jariwala[1]*

[1]Electrical and Systems Engineering, University of Pennsylvania, Philadelphia, PA, USA
[2]Materials Science and Engineering, University of Pennsylvania, Philadelphia, PA, USA
[3]Inter-university Microelectronics Center (IMEC), Leuven, Belgium
[4]Department of Physics and Astronomy University of Missouri, Columbia, USA
[5]National Institute of Standards and Technology, Gaithersburg, MD, USA
[6]Low-dimensional Semiconductors Lab, Metallurgical and Materials Engineering, Indian Institute of Technology-Madras, Tamilnadu, India
[7]Argonne National Laboratory, Illinois, USA
[8]Center for Nanophase Materials Sciences, Oak Ridge National Laboratory, Oakridge, USA
[9]Materials Science & Engineering, University of Washington, Seattle, USA
[10]Department of Mechanical Engineering and Texas Materials Institute, University of Texas, Austin, USA

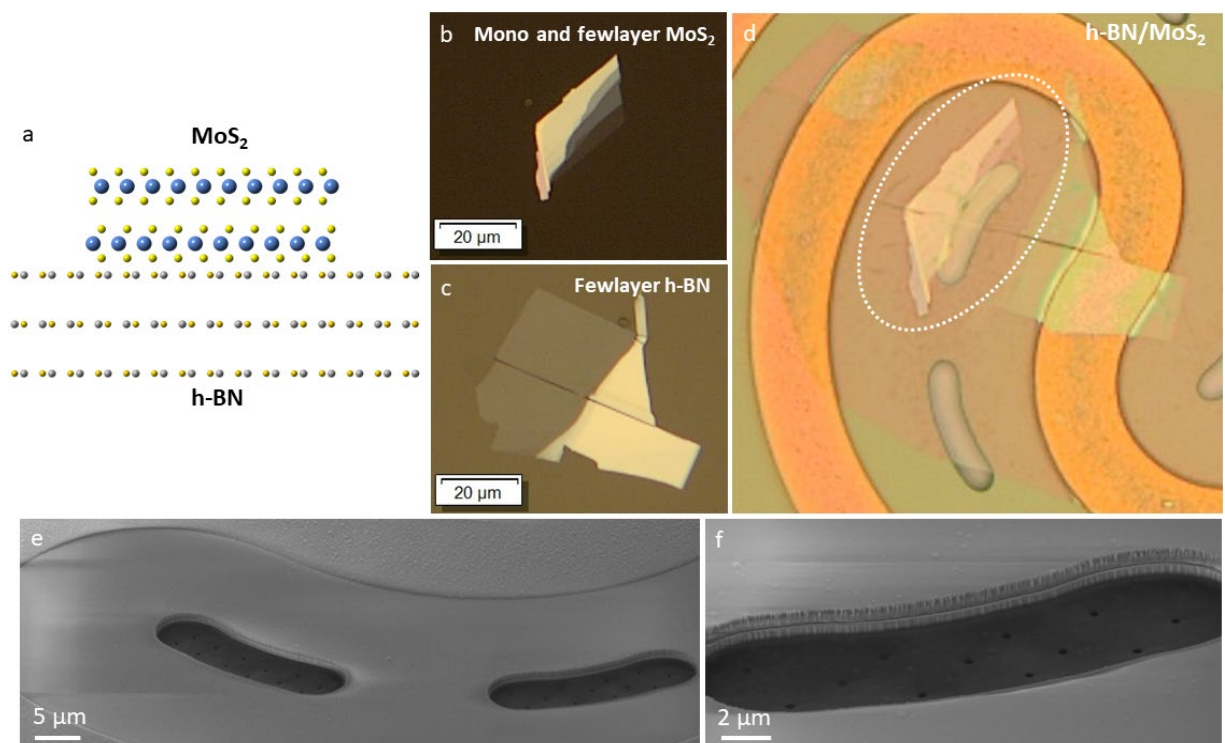

Figure S1: (a) An atomic model shows the stack configuration of non-encapsulated condition and corresponding (b, c) Optical images of individual 2D layers on Polydimethylsiloxane (PDMS) stamp which were exfoliated and isolated from respective bulk crystals, (d) An optical micrograph of the region where site-specific transfer of h-BN and $MoS_2$ layers was done one after other along-with forming gas annealing, (e, f) Low and high magnification scanning

electron microscopic image of the TEM transparent region in Micro-electromechanical systems (MEMS) based heating chip where sub-micron holes are created in $SiN_x$ membrane using $Xe^+$ plasma-based focus ion beam (FIB) technique.

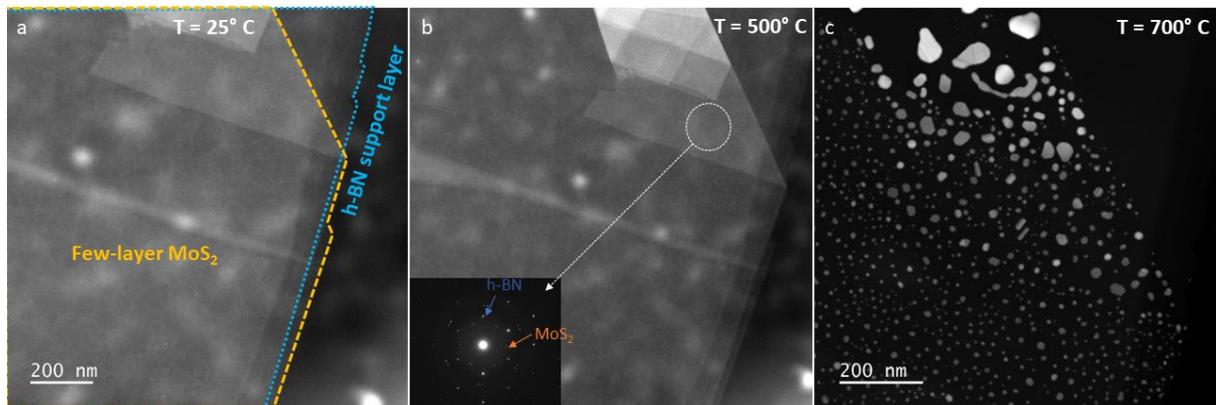

Figure S2: Dark-field STEM image of h-BN /$MoS_2$ stack at (a) room temperature and (b) at 500 °C which do not show any significant changes. Inset shows the fast-Fourier transform (FFT) pattern reflecting the $MoS_2$ as well as h-BN crystal pattern. (c) $MoS_2$ layer undergoes thermolysis at 700 °C, showing the disintegration into tiny islands.

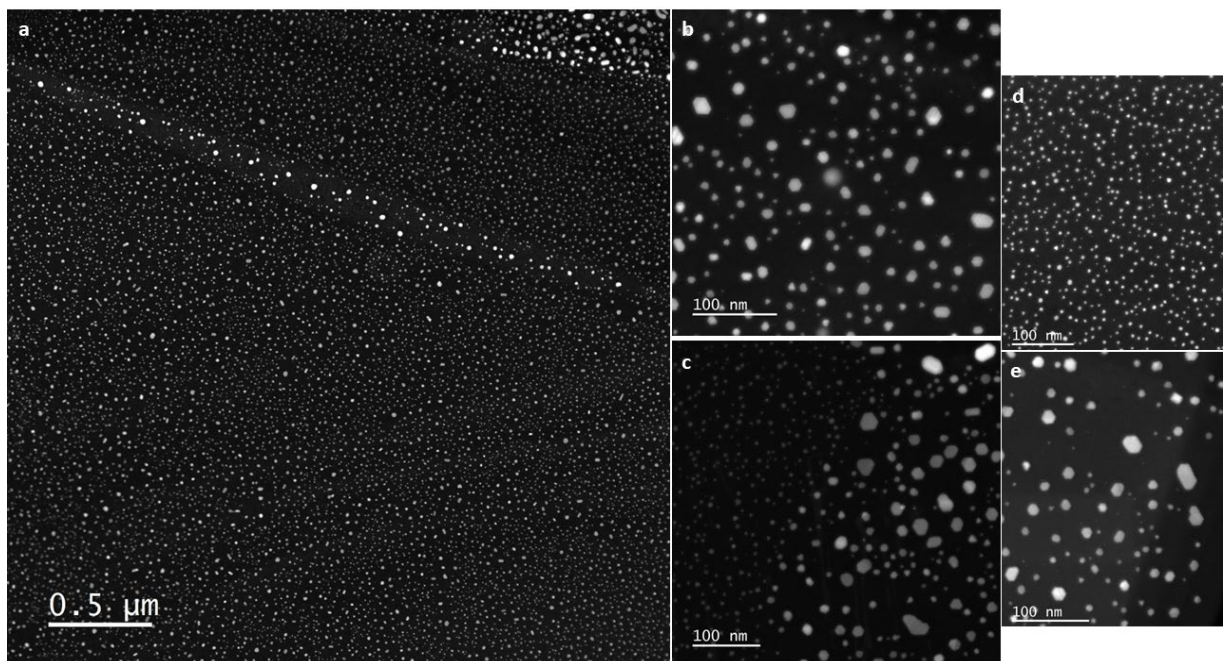

Figure S3: Dark-field scanning transmission electron microscopy (STEM) images of a uniform and controlled nanoparticle formation based upon the thicknesses of the parent transition metal dichalcogenides (TMDCs) $MoS_2$ layer. After thermolysis, we can also see the wrinkled region transformed into thicker particles (brighter contrast) shown in (a).

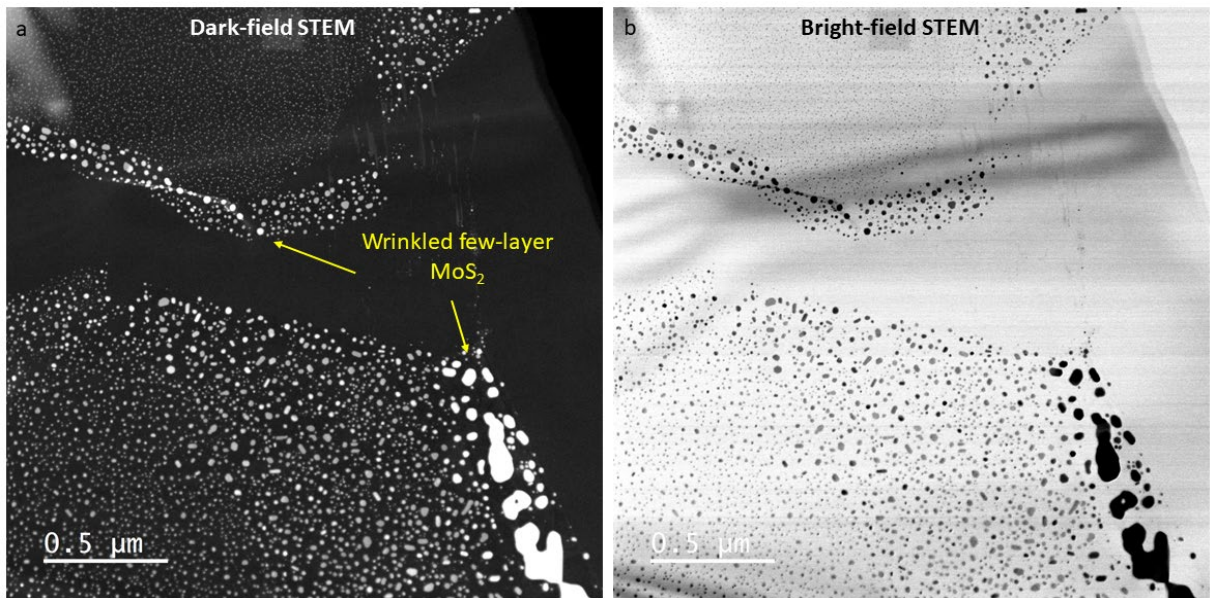

Figure S4: Dark-field and bright-field STEM images are presented side by side respectively for same region of sample after non-equilibrium thermolysis of MoS$_2$ where we can see the disintegrated MoS$_2$ particles formed along-with supported bottom layer of h-BN.

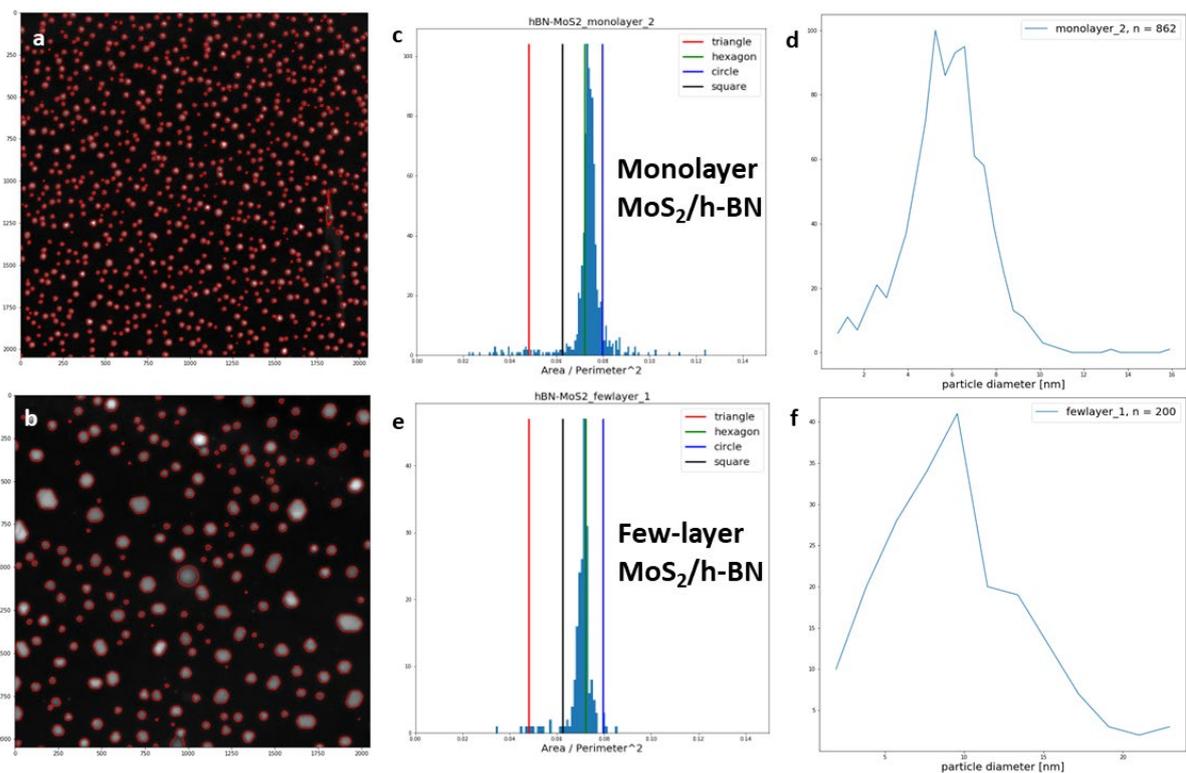

Figure S5: (a, b) Convolutional Neural Network (CNN) based machine learning adopted to identify selectively the disintegrated particles as shown and corresponding shape-size distribution. (c, d) Shape of MoS$_2$ particles from monolayer disintegration which mostly attaining the shape of circular to hexagon while the average particle size comes around (5 to 6) nm. Similarly, (e, f) Shape of MoS$_2$ particles from few-layer disintegration which mostly attain the shape of hexagon while the average particle size comes around 10 nm.

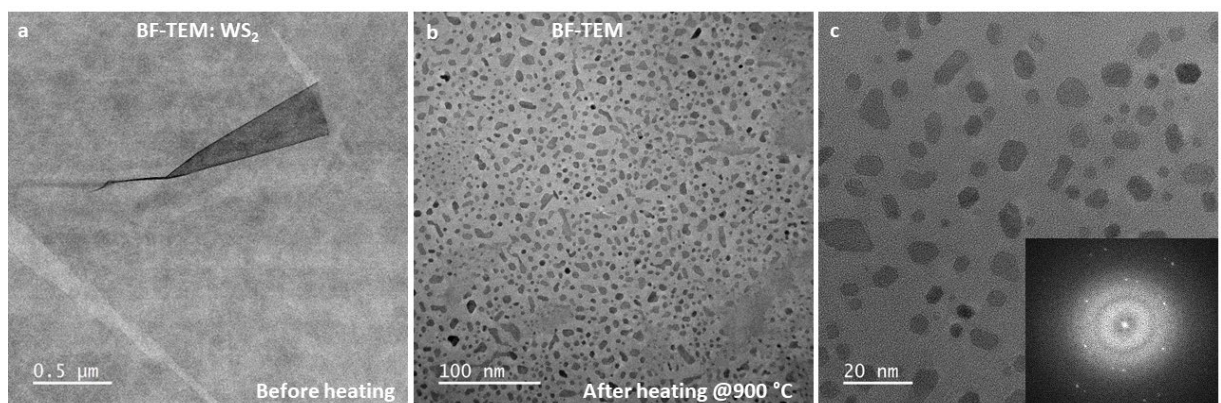

Figure S6: Bright-field TEM image of monolayer WS$_2$ (a) before heating and (b) after heating at 900 °C. (c) High-resolution TEM image of as formed WS$_2$ particles and added FFT pattern as inset.

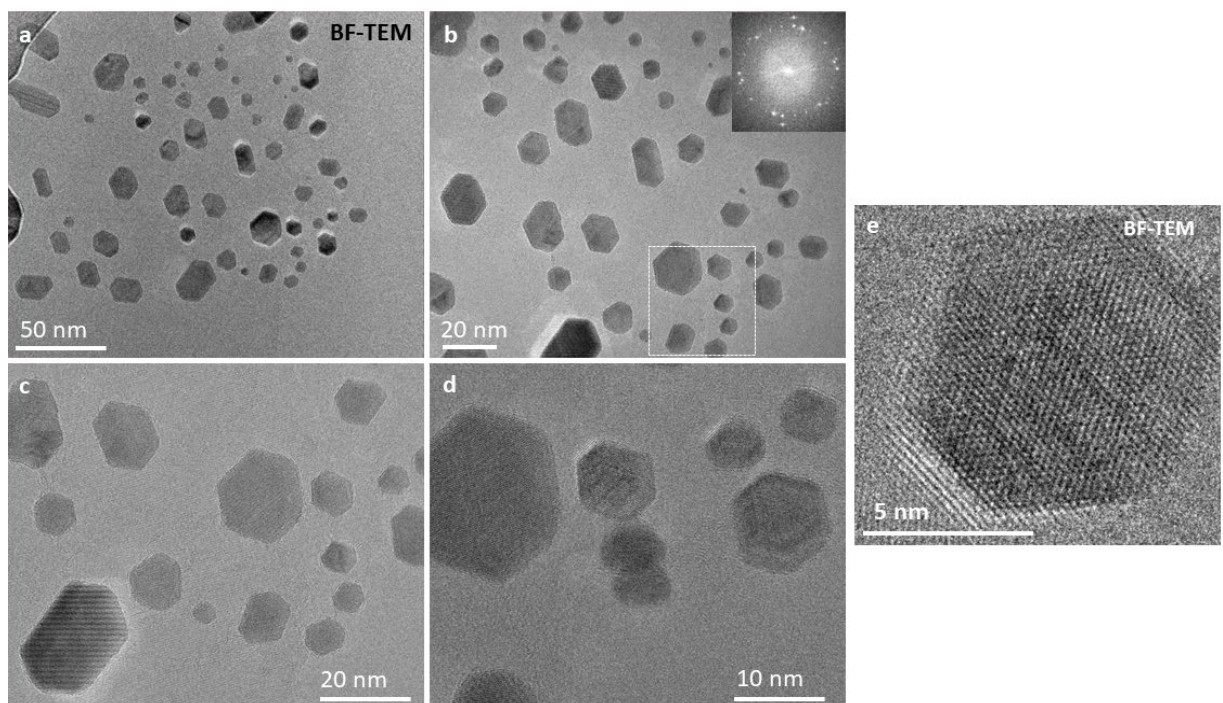

Figure S7: High-resolution TEM images of hexagonal nanoparticles after thermolysis of MoS$_2$ layers.

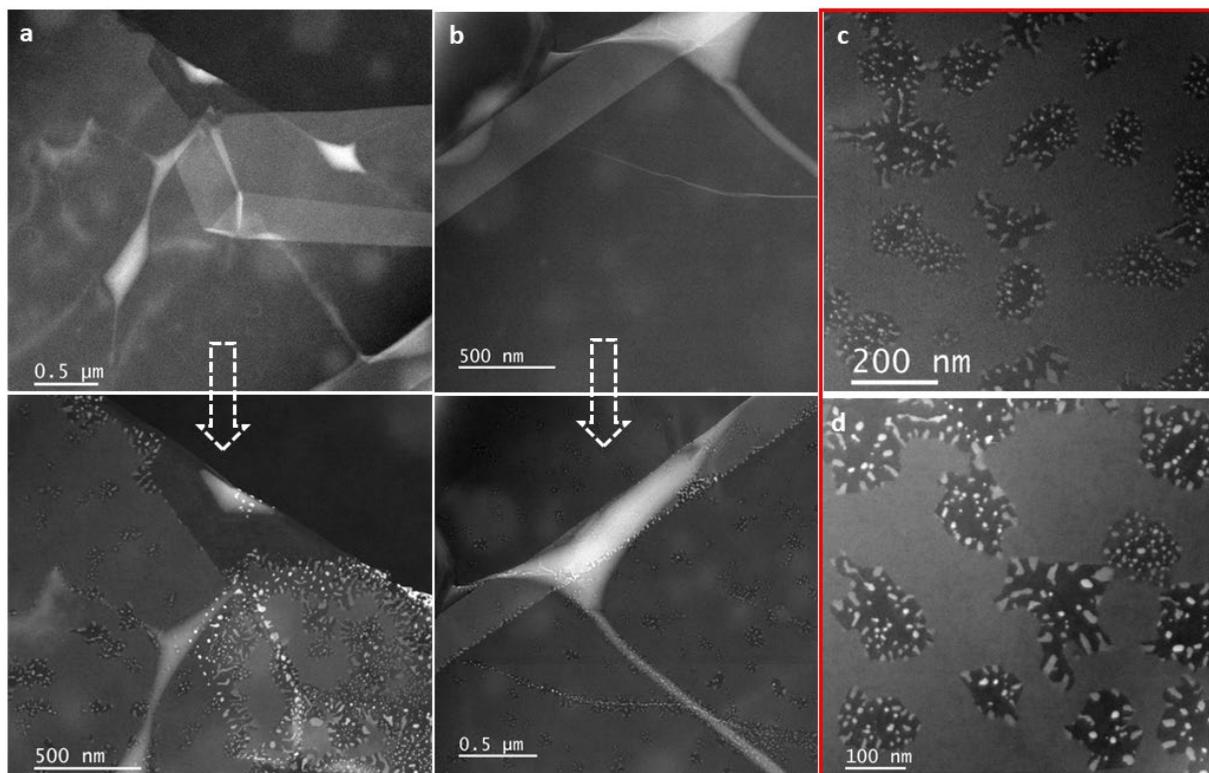

Figure S8: (a, b) A non-uniform disintegration after thermolysis due to the presence of leftover PDMS residues (easily recognised through blurred bright contrasts in the preheated case) in between h-BN and $MoS_2$ layers. (c) Magnified STEM images show the $MoS_2$ nanoparticles formation in pockets of disintegrated regions.

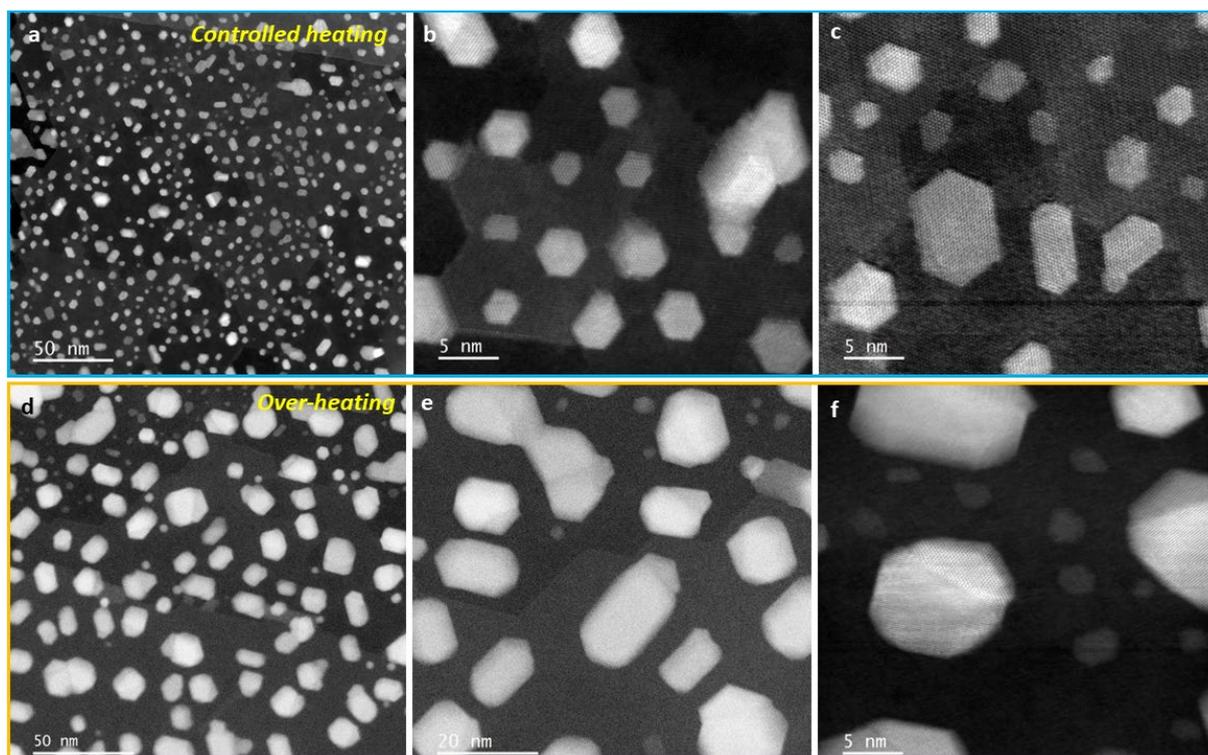

Figure S9: (a-c) Thermolysis (controlled heating) of few-layer MoS$_2$ shows the layer-by-layer transformation to nanoparticles, also easily recognise the left-behind intact continuous MoS$_2$ layer, (d-f) Similar few-layer MoS$_2$ undergone to the thermolysis process for a relatively long period (over-heating) to make sure the complete layer transformation, which results to formation of mostly thicker Mo nanocrystals (brighter contrasts).

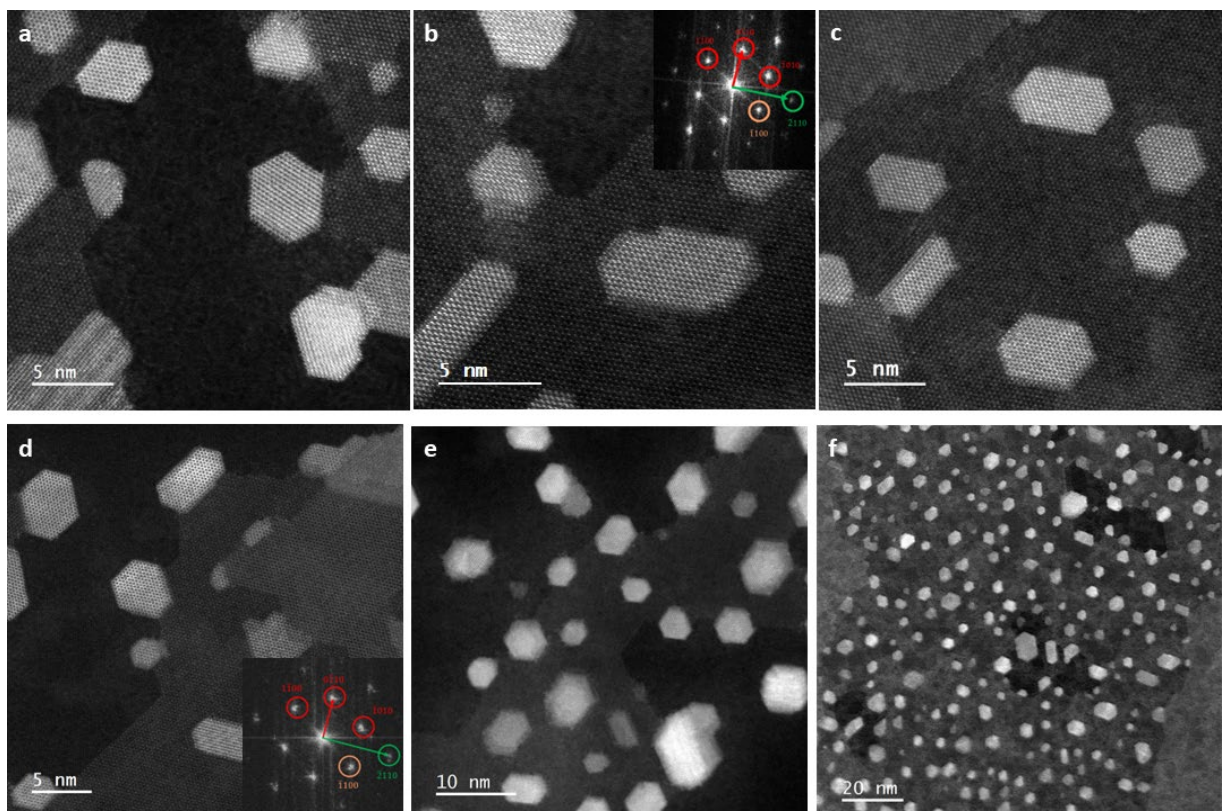

Figure S10: Atomically resolved high-angle annular dark field (HAADF)-STEM images show the various stages of transformation during thermolysis in the case of bilayer/few-layer $MoS_2$. We can corroborate from the multiple images that the diffusion mostly started from the top layer and open edges (the rate is faster at the edges) and then goes to the bottom forming tiny islands of $MoS_2$.

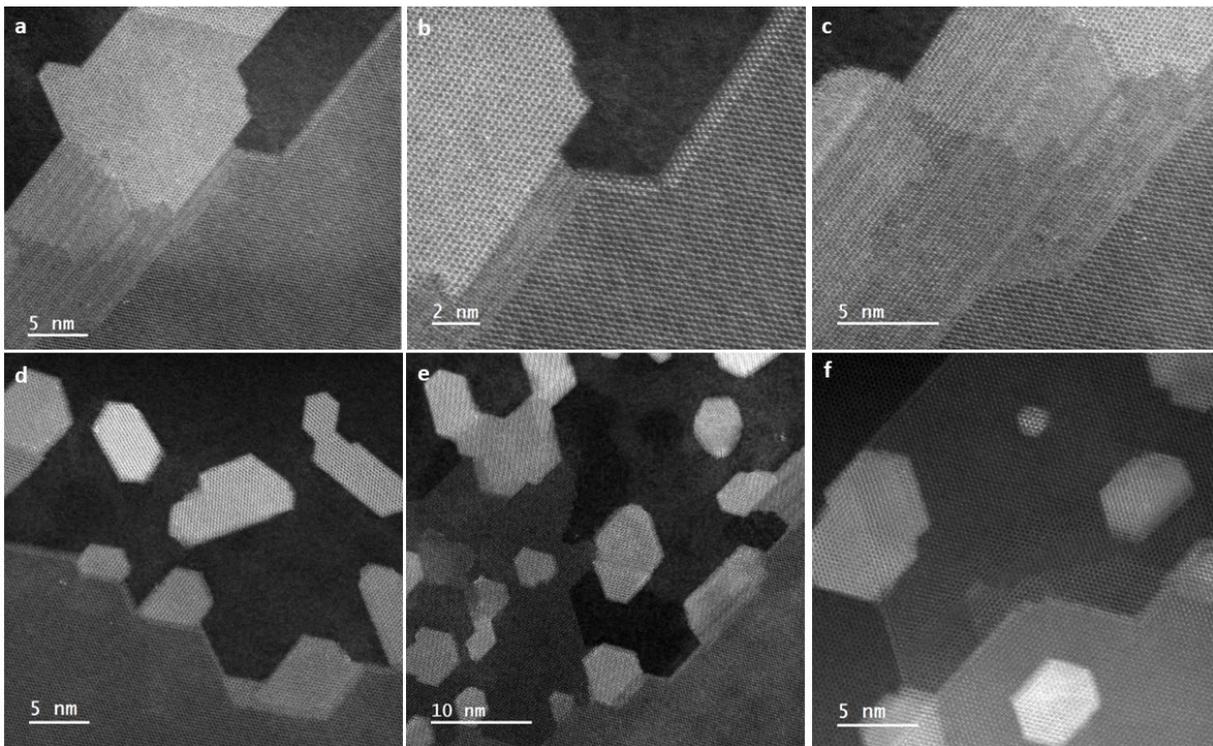

Figure S11: Restructuring of the edges during the conversion of $MoS_2$ layers into particles when the rate of transformation is relatively smaller than usual.

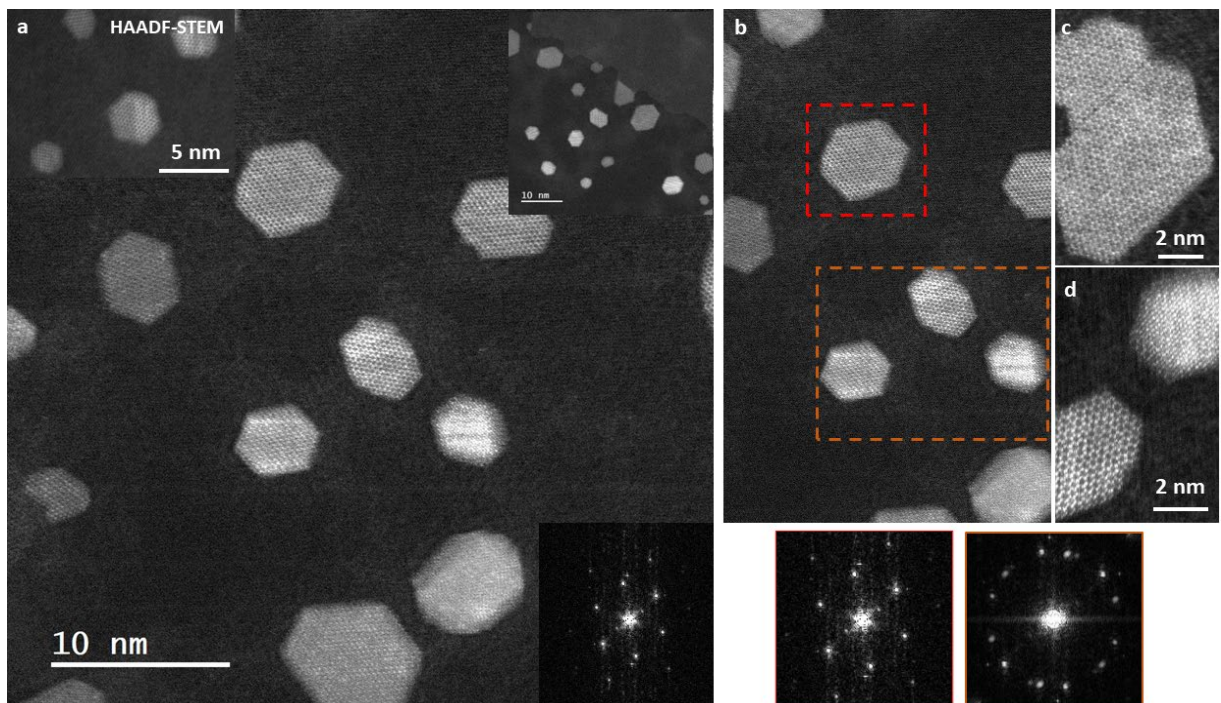

Figure S12: (a-d) Ultra-small $MoS_2$ particle formation after thermolysis and their orientation variation relative to the neighbouring particles as seen in the attached FFT pattern (red and orange rectangular regions).

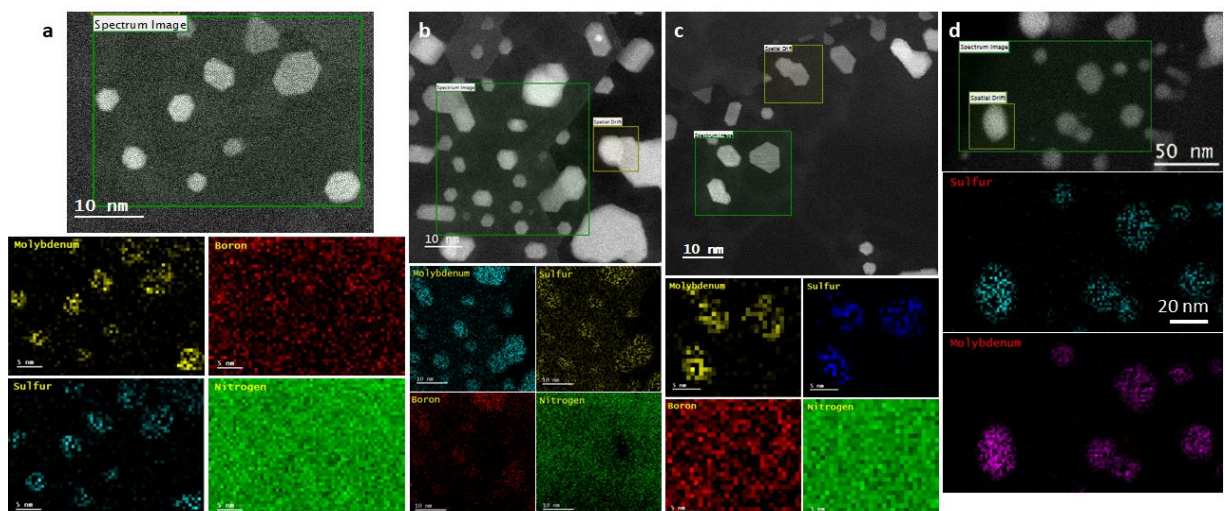

Figure S13: EDS mapping for the various regions (from different samples) shows the formation of MoS$_2$ particles with extremely confined dimensions ranging from (2 to 5) nm only.

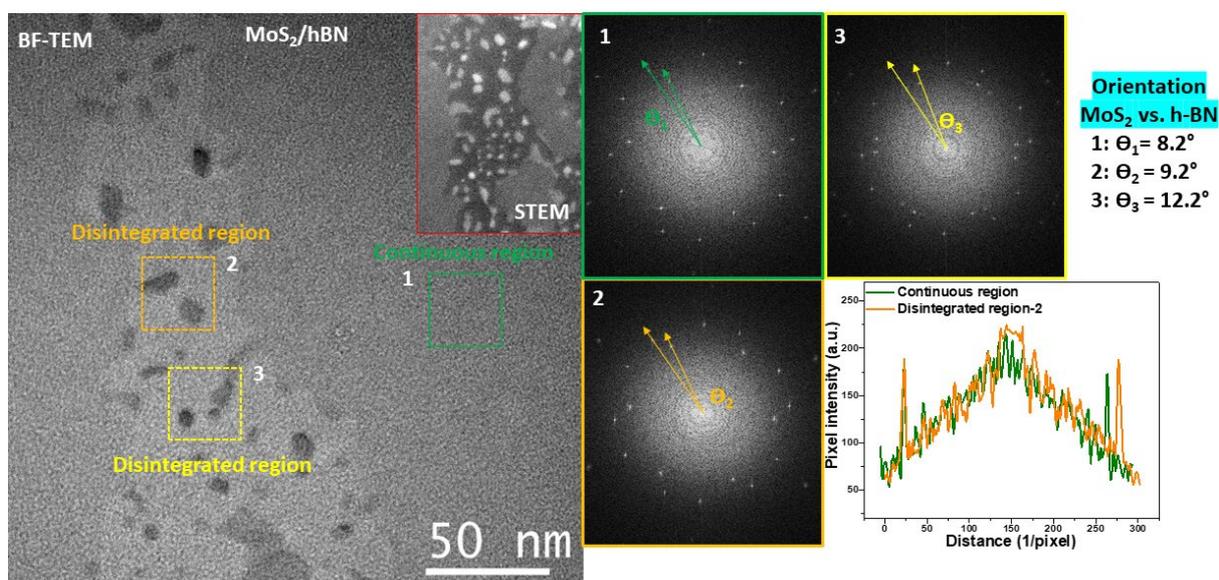

Figure S14: Disintegrated nanoparticles are analysed with corresponding FFT patterns (1, 2 and 3) which can be seen and compared from different spots, indicating the randomness of orientation relative to the h-BN support layer after thermolysis.

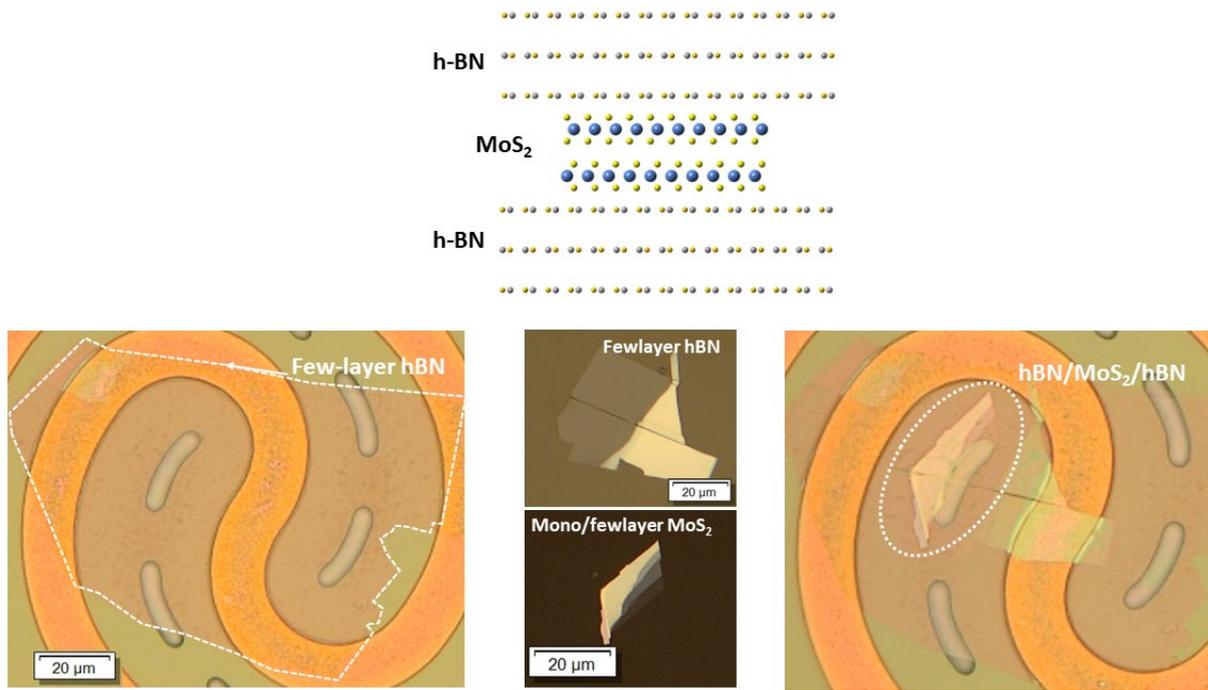

Figure S15: Atomic model showing the layer configuration for encapsulated hBN-$MoS_2$-hBN hetero system and corresponding the transferred few-layer h-BN and $MoS_2$ onto the MEMS-based prefabricated heating chip at a dedicated location (e-transparent bean-shape region) to ensure the TEM/STEM imaging while we perform in-situ thermolysis.

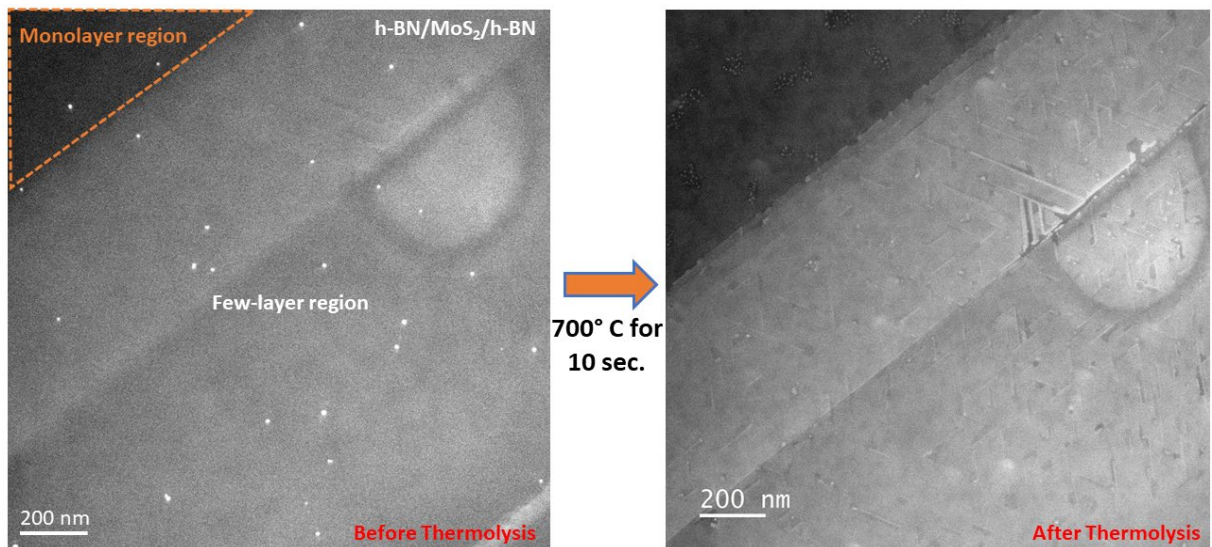

Figure S16: Low-magnification STEM images of the encapsulated $MoS_2$ before and after the thermolysis, processed for 10 sec at 700 °C.

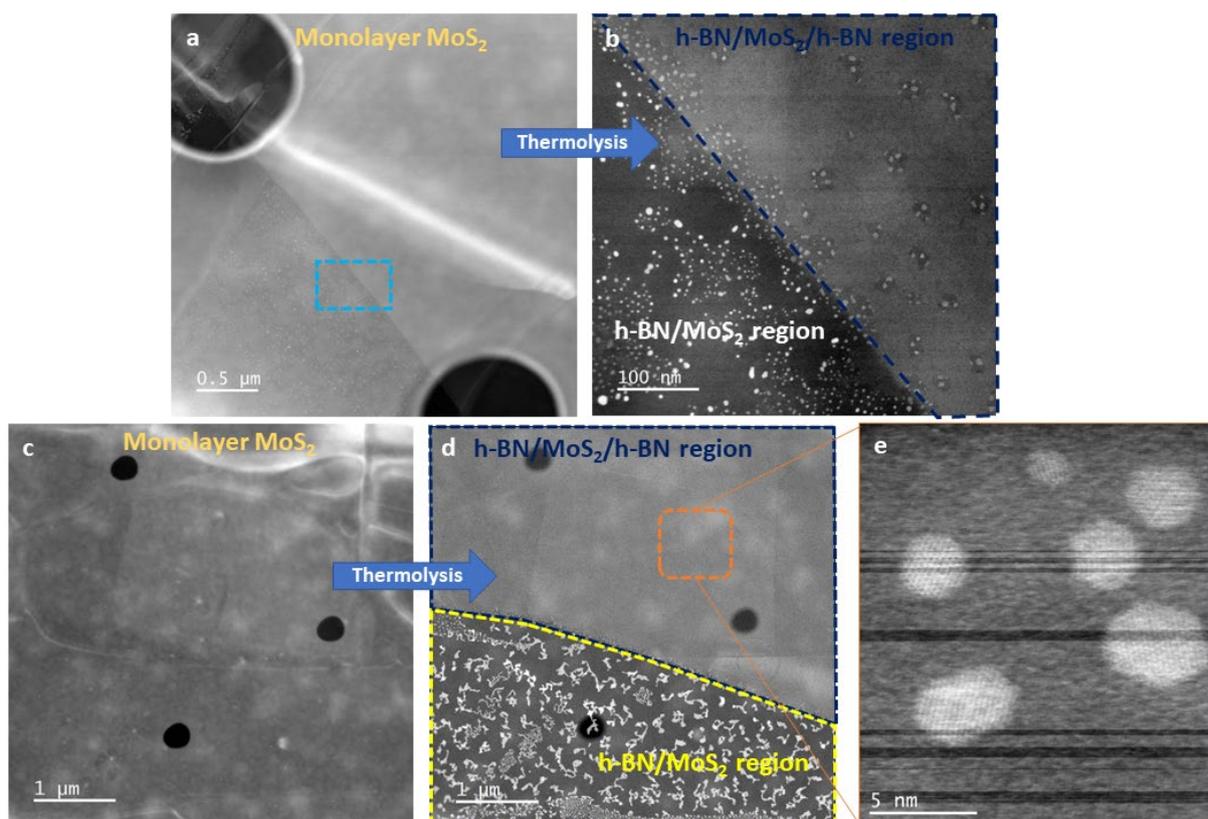

Figure S17: STEM imaging of half-encapsulate monolayer MoS$_2$ layers which is captured before and after (Fig a Vs b and Fig c Vs d) the thermolysis process. It clearly visualises the impact of the encapsulated regions where thermolysis occurred relatively slower than the non-encapsulated (h-BN/MoS$_2$ only) region. In fig. a and b, MoS$_2$ inside the encapsulated region is partly disintegrated while the non-encapsulated one undergoes a complete transformation. Similarly, when we transformed completely the encapsulated MoS$_2$ (as shown in Fig e) at the same time non-encapsulated MoS$_2$ transformed into metal particles due to overheating in this case.

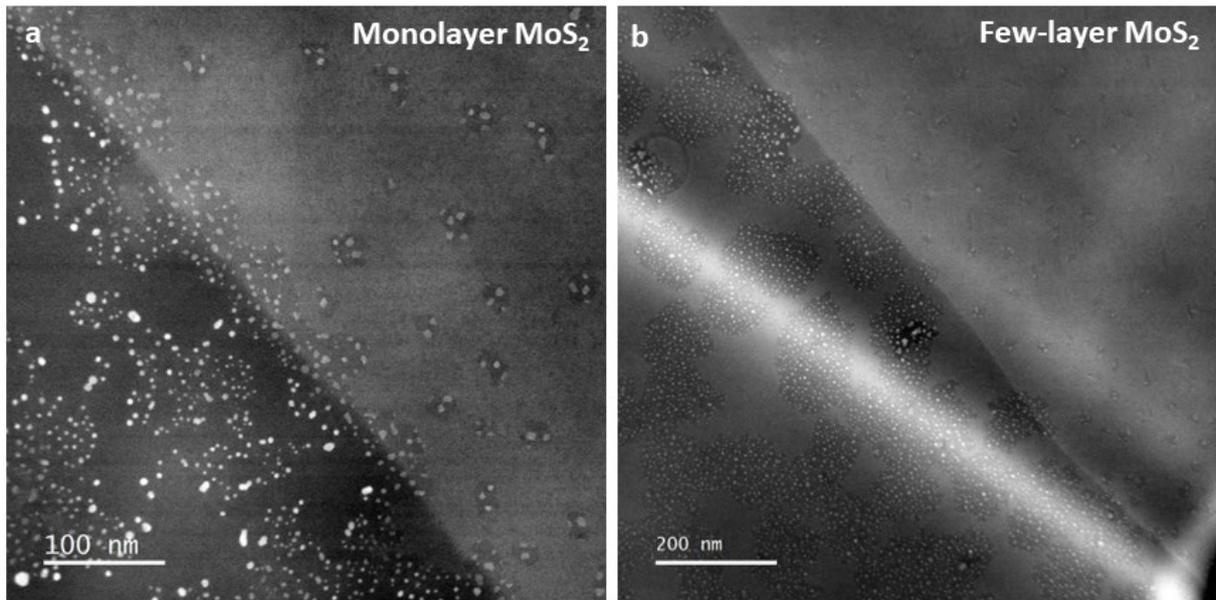

Figure S18: Dark-field STEM images of the transformed monolayer and few-layer MoS$_2$ in the partly encapsulated as well as partly non-encapsulated configurations.

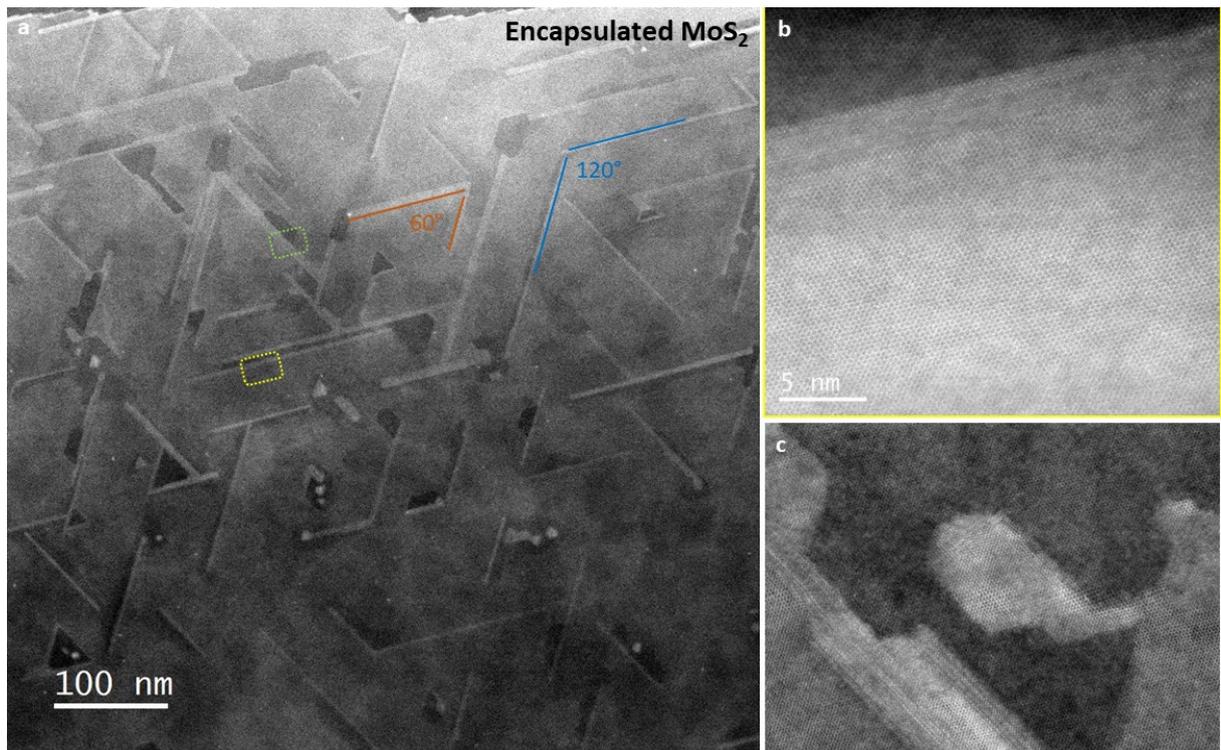

Figure S19: (a) Faceting of the trenches under encapsulated configuration reveals the orientation with the MoS$_2$ plane direction, (b, c) Atomically resolved STEM images show the nature of the MoS$_2$ phase (2H) in the formation of the trench as well as a remaining bottom layer of MoS$_2$.

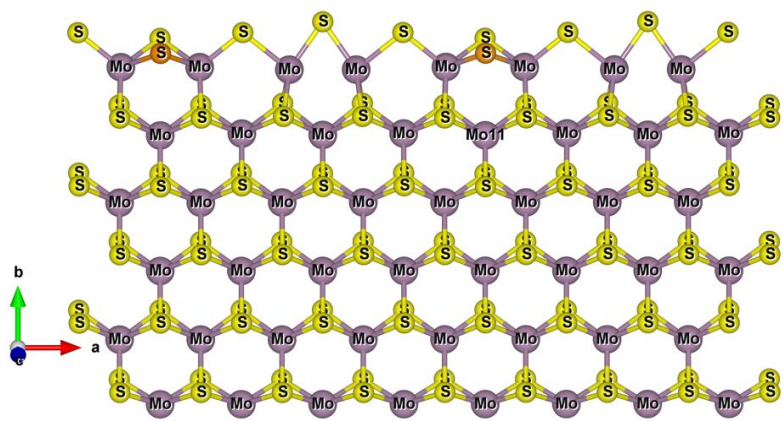

Figure S20: MoS$_2$ nanoribbon with sulfur vacancies modelled at the edge.

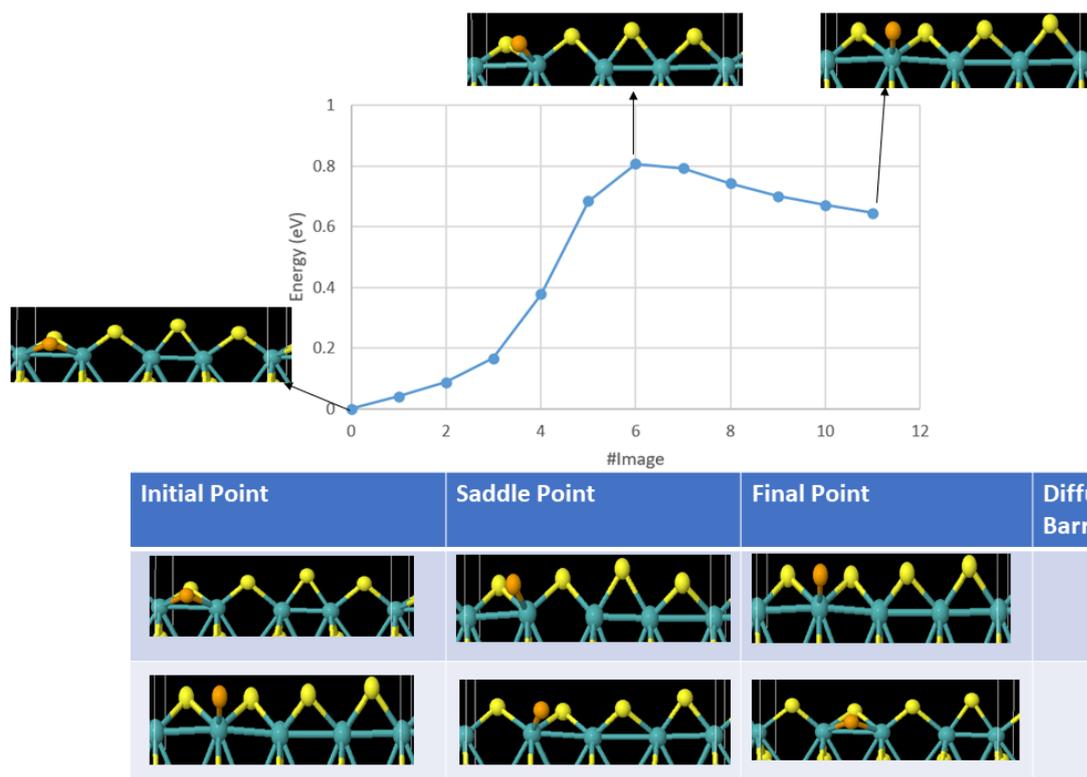

Figure S21: Figure demonstrating sulfur diffusion at the edges in MoS$_2$ structure computed using density functional theory (DFT) + nudged elastic band (NEB) methodology between initial and intermediate state. It has been shown that the diffusion along the edge is feasible/fast at higher temperature with an energy barrier of 0.8 eV. The rate of diffusion is faster than the rate of evaporation under confinement leading to a thermodynamically equilibrium shape of triangle.

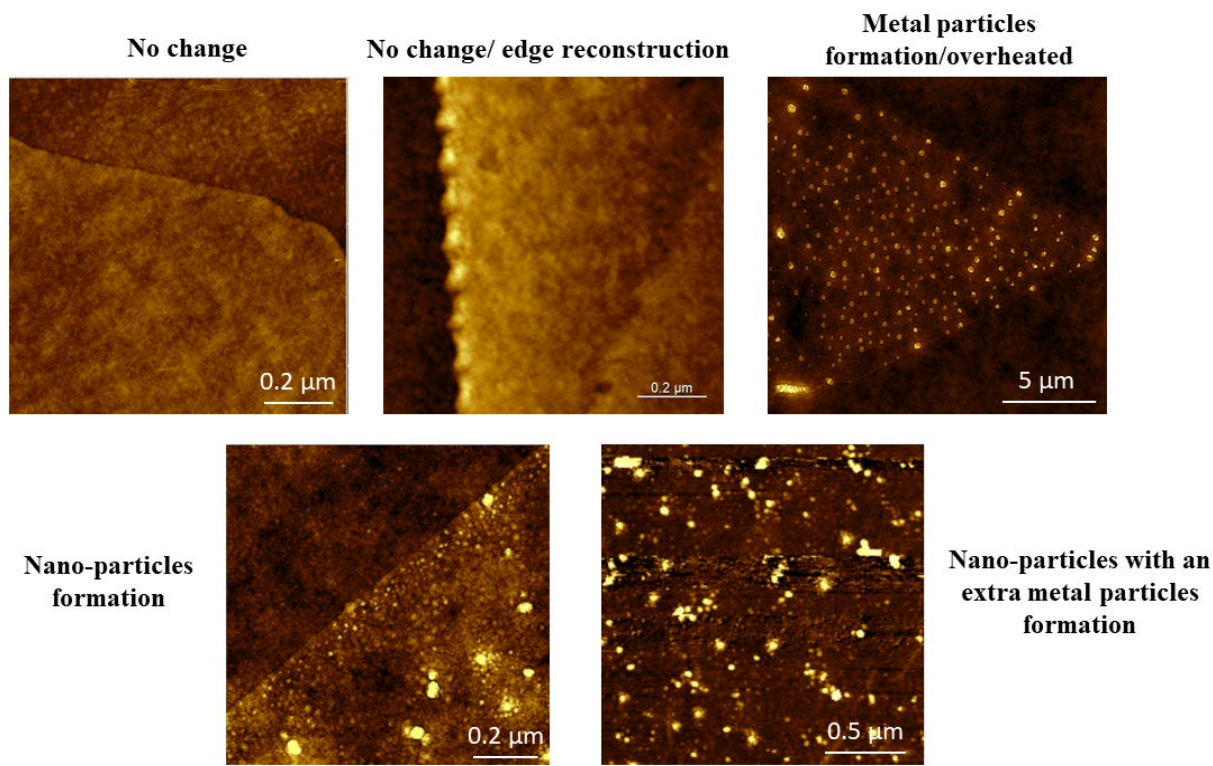

Figure S22: Atomic force microscopic (AFM) height images (MoS$_2$ layers placed on SiO$_2$/Si substrate) were taken at different stages during the optimisation of MoS$_2$ nano-particles formation using ex-situ processing.

| Target Temperature (Ex-situ processing) | Formation Time (Seconds) | Formation Results (Disintegration of MoS$_2$ continuous layer) |
|---|---|---|
| 800°C | 15s | No change |
| 800°C | 30s | No change/edge change |
| 800°C | 45s | Nano-particles formation |
| 800°C | 60s | Metal particles formation |
| 825°C | 25s | No change/A very little changes |
| 825°C | 35s | Metal particles formation |
| 850°C | 15s | Nano-particles formation with small metal clusters |
| 850°C | 25s | Metal particles formation |

TABLE S1: Optimisation parameters to achieve disintegrated nanoparticles of MoS$_2$ for the ex-situ thermolysis process carried out with the help of a modified quartz-tube furnace.

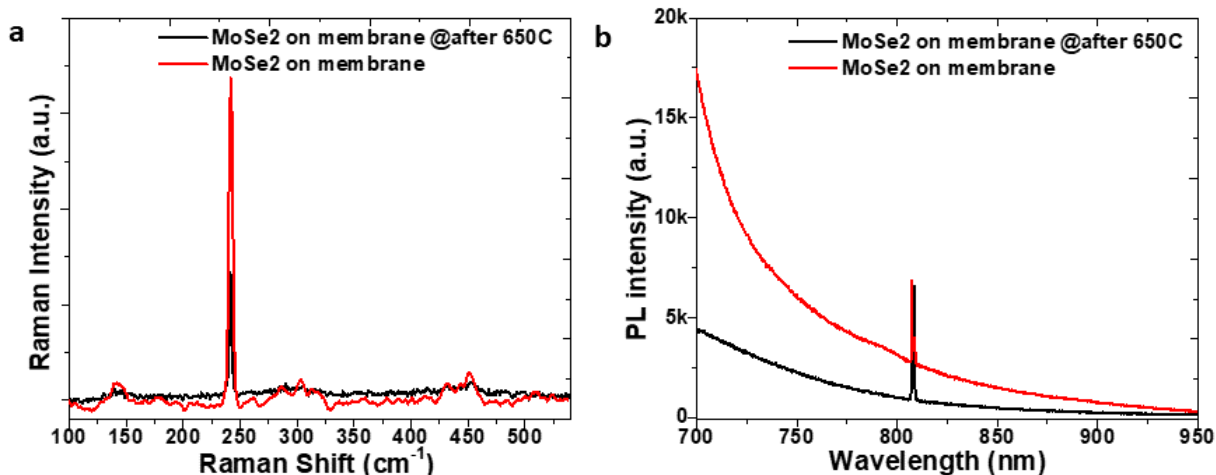

Figure S23: Near-field tip-enhanced Raman and PL (TERS and TEPL) spectroscopic analysis for the MoSe$_2$ layers before and after the thermolysis at 650 °C measured directly at the membrane region of the MEMS-TEM grid.

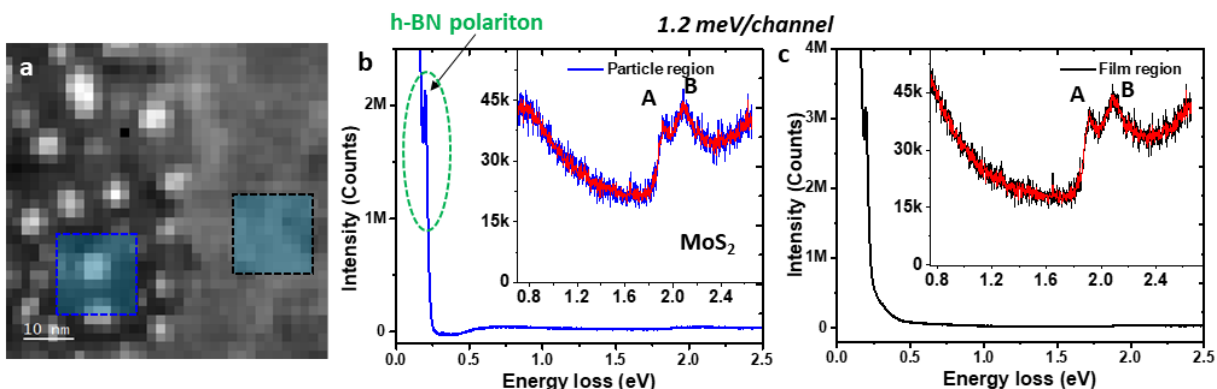

Figure S24: HAADF-STEM image selected for the analysis of the band-edge electron energy loss (EELS) after the thermolysis of the few-layer MoS$_2$. Figure b and c show the EELS spectra from the two different regions as mentioned. The h-BN polariton peak position (as marked in fig b) is used for the referencing of A and B-exciton positions obtained for MoS$_2$ while spectra are measured with a very small channel window of 1.2 meV only.

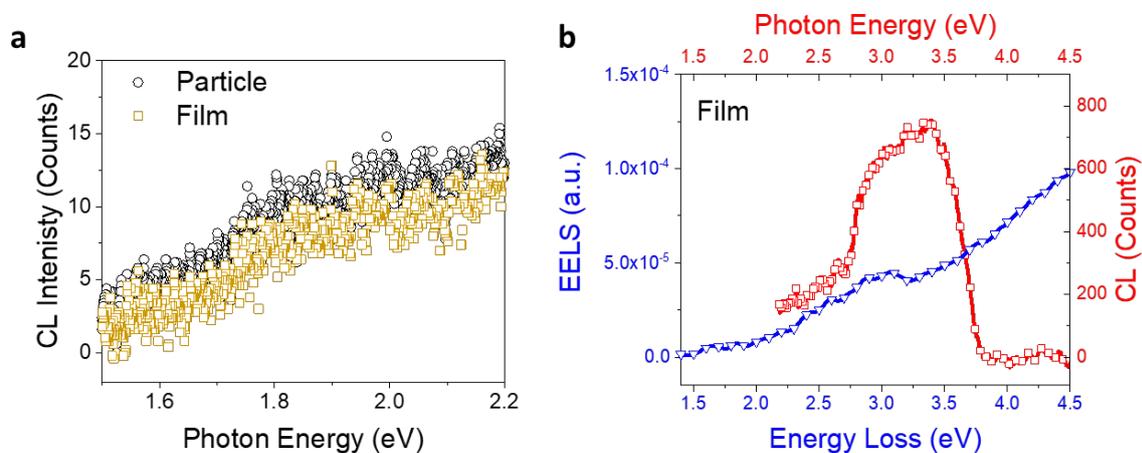

Figure S25: Cathodoluminescence (CL) emissions after rapid thermolysis. (a) CL spectra in the energy range of the direct band edge from the two different regions: disintegrated particle

(labelled as Particle) and continuous layer with overlapped particle (labelled as Film). (b) CL spectrum of the film region showing the prominent C-exciton emission.

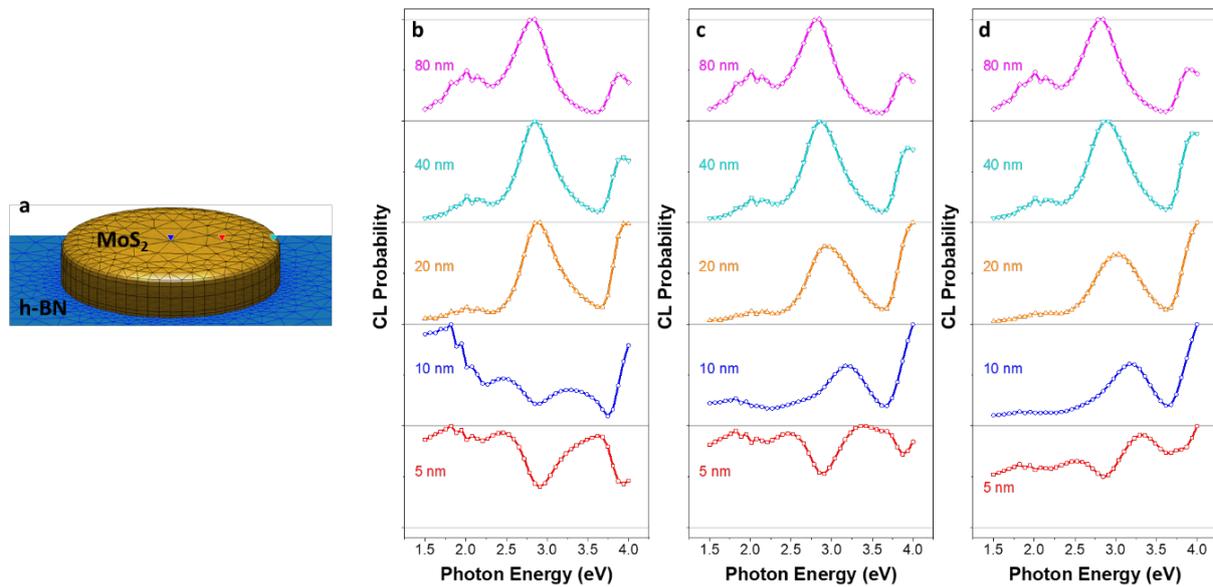

Figure S26. Theoretical prediction of CL emission from a $MoS_2$ disk support by a h-BN substrate using the MNPBEM toolbox described in Ref. 31 in the main text. Three-dimensional model of discretized interfaces that separate the vacuum (refractive index, n ≈ 1), the 1-nm-thick $MoS_2$ disk with a diameter of $D$ (complex refractive index[1], and the **1-nm-thick** h-BN (n ≈ 2.13). Electron beam impact positions are indicated and color-coded at the center (blue), 0.24*$D$ away from the center (red), and 0.48*$D$ away from the center (turquoise). Calculated CL probability for $D$ = 80 nm (magenta, rhombus), 40 nm (turquoise, downward triangle), 20 nm (orange, upward triangle), 10 nm (blue, circle), and 5 nm (red, square) with an electron impact position at the (b) center, (c) middle (0.24*D away from the center), (d) edge (0.48*D away from the center).